\PassOptionsToPackage{dvipsnames}{xcolor}
\documentclass[11pt,letterpaper]{swepreprint}

\usepackage[numbers,sort&compress]{natbib}
\usepackage{graphicx}
\usepackage{amsmath,amsfonts,amssymb}
\usepackage{booktabs}
\usepackage{tabularx}
\usepackage[table]{xcolor}
\usepackage{multirow}
\usepackage{caption}
\usepackage{wrapfig}
\usepackage{float}
\usepackage{xspace}
\usepackage{pifont}
\usepackage{fontawesome5}
\usepackage{needspace}
\usepackage{placeins}
\usepackage{listings}
\tcbuselibrary{listings,skins,breakable}
\hypersetup{
  pdftitle={SWE-Touch: Benchmarking Coding Agents When Users Touch the Code},
  pdfauthor={Yuntao Tan, Yige Song, Jiacheng Guo, Yuxin Ren, Yining Ye, Xingyao Wang, Yujia Qin, Zhiyuan Liu, Maosong Sun},
  pdfsubject={Software Engineering, AI Agents, Benchmark},
  pdfkeywords={coding agents, user interaction, benchmark, SWE-bench}
}
\newcommand{\method}{\textsc{SWE-Touch}\xspace}
\newcommand{\vanilla}{\textsc{Vanilla}\xspace}
\newcommand{\counteredit}{\textsc{Counter-Edit}\xspace}
\newcommand{\coedit}{\textsc{Co-Edit}\xspace}
\newcommand{\featureyes}{\textcolor{green!55!black}{\ding{51}}}
\newcommand{\featureno}{\textcolor{red!75!black}{\ding{55}}}
\newcommand{\dropvalue}[1]{#1}
\newcommand{\risevalue}[1]{#1}
\newcommand{\modellogo}[1]{\raisebox{-0.1ex}{\includegraphics[height=1.55ex,width=1.75ex,keepaspectratio]{Figures/logo_#1.png}}\hspace{0.28em}}
\newcommand{\rank}[1]{\,{\scriptsize(\textbf{#1})}}
\newcolumntype{Y}{>{\raggedright\arraybackslash}X}

\newtcolorbox{promptmodule}[2][]{
  enhanced,colback=BGBlue!70!white,colframe=ThemeBorder,
  colbacktitle=ThemeBorder!18!BGBlue,coltitle=TitleText,fonttitle=\bfseries,
  title={#2},boxrule=0.7pt,arc=0pt,left=7pt,right=7pt,top=5pt,bottom=5pt,#1}
\newtcolorbox{interactioncase}[2][]{
  enhanced,colback=BGBlue!65!white,colframe=ThemeBorder,
  colbacktitle=ThemeBorder!16!BGBlue,coltitle=TitleText,fonttitle=\bfseries,
  title={#2},title after break={#2 (continued)},boxrule=0.7pt,arc=0pt,
  left=7pt,right=7pt,top=5pt,bottom=5pt,#1}
\newtcblisting{tracecode}[1]{
  enhanced,breakable,listing only,width=\linewidth,
  colback=black!2,colframe=black!28,colbacktitle=black!8,
  coltitle=black,fonttitle=\bfseries,title={#1},
  title after break={#1 (continued)},boxrule=0.45pt,arc=1pt,
  left=5pt,right=5pt,top=3pt,bottom=3pt,
  listing options={language={},basicstyle=\ttfamily\scriptsize,
    keywordstyle=\color{black},commentstyle=\color{black},
    stringstyle=\color{black},identifierstyle=\color{black},
    backgroundcolor=\color{black!2},
    numbers=left,numberstyle=\tiny\color{black!45},
    numbersep=0.65em,stepnumber=1,numberblanklines=true,
    breaklines=true,breakatwhitespace=false,columns=fullflexible,
    keepspaces=true,showstringspaces=false,frame=none,framerule=0pt,
    resetmargins=true,xleftmargin=2.2em,xrightmargin=0pt,
    framexleftmargin=0pt,framexrightmargin=0pt,
    aboveskip=0pt,belowskip=0pt}}
\newtcblisting{judgeprompt}[1]{
  enhanced,breakable,listing only,width=\linewidth,
  colback=BGBlue!68!white,colframe=ThemeBorder,colbacktitle=ThemeBorder!20!BGBlue,
  coltitle=TitleText,fonttitle=\bfseries,title={#1},
  title after break={#1 (continued)},boxrule=0.55pt,arc=1pt,
  left=5pt,right=5pt,top=4pt,bottom=4pt,
  listing options={language={},basicstyle=\ttfamily\footnotesize\color{black},
    keywordstyle=\color{black},commentstyle=\color{black},
    stringstyle=\color{black},identifierstyle=\color{black},
    backgroundcolor=\color{BGBlue!68!white},
    breaklines=true,breakatwhitespace=true,columns=fullflexible,
    keepspaces=true,showstringspaces=false,numbers=none,
    frame=none,framerule=0pt,
    resetmargins=true,xleftmargin=0pt,xrightmargin=0pt,
    framexleftmargin=0pt,framexrightmargin=0pt,
    aboveskip=0pt,belowskip=0pt}}
\newtcolorbox{traceround}[1]{
  enhanced,breakable,width=\linewidth,colback=white,colframe=black!22,
  colbacktitle=black!7,coltitle=black,fonttitle=\bfseries,
  title={#1},title after break={#1 (continued)},boxrule=0.45pt,arc=1pt,
  left=6pt,right=6pt,top=4pt,bottom=4pt}
\newtcolorbox{traceoutcome}{
  enhanced,breakable,colback=BGBlue!60!white,colframe=ThemeBorder,
  boxrule=0.45pt,arc=1pt,left=6pt,right=6pt,top=4pt,bottom=4pt}

\newcommand{\TitleSubtitleFont}{\fontsize{16.5}{19}\selectfont} 
\title{\texorpdfstring{%
\raisebox{-0.2\height}{\includegraphics[height=2.5em]{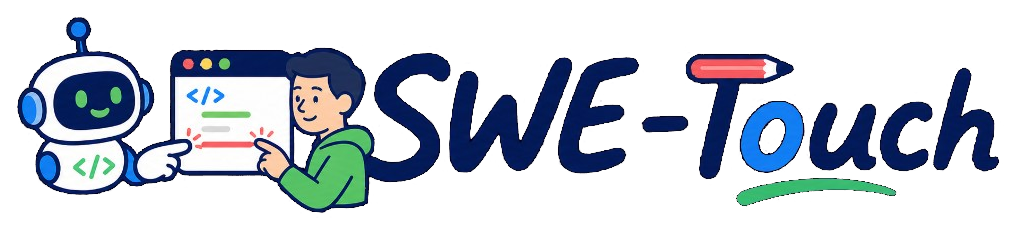}}%
\hspace{0.18em} \\ {\TitleSubtitleFont Benchmarking Coding Agents When Users Touch the Code}%
}{SWE-Touch: Benchmarking Coding Agents When Users Touch the Code}}
\runningtitle{SWE-Touch: Benchmarking Coding Agents When Users Touch the Code}
\newcommand{\TitleLinks}{%
  \vspace{6pt}
  {\centering\small
  \href{https://github.com/Trae1ounG/SWE-Touch}{\faGithub}\hspace{0.2em}
  \textbf{Project: }\href{https://github.com/Trae1ounG/SWE-Touch}
  {\texttt{github.com/Trae1ounG/SWE-Touch}}\par
  \vspace{2pt}
  \href{https://huggingface.co/datasets/Trae1ounG/SWE-Touch}{\faDatabase}\hspace{0.2em}
  \textbf{Dataset: }\href{https://huggingface.co/datasets/Trae1ounG/SWE-Touch}
  {\texttt{huggingface.co/datasets/Trae1ounG/SWE-Touch}}\par}}
\newcommand{\authmark}[1]{{\mdseries\textsuperscript{#1}}}
\author{%
  {\Authfont
  \textbf{Yuqiao Tan}\authmark{1,2}\quad
  \textbf{Jinxiang Meng}\authmark{1,2}\quad
  \textbf{Fangyu Lei}\authmark{1,2}\quad
  \textbf{Minzheng Wang}\authmark{1,2}}\\
  {\Authfont
  \textbf{Shizhu He}\authmark{1,2,\textdagger}\quad
  \textbf{Jun Zhao}\authmark{1,2}\quad
  \textbf{Kang Liu}\authmark{1,2}}\\
  {\Affilfont
  \textsuperscript{1}Institute of Automation, Chinese Academy of Sciences\quad
  \textsuperscript{2}University of Chinese Academy of Sciences\\
  \texttt{tanyuqiao2025@ia.ac.cn}\quad
  \texttt{shizhu.he@nlpr.ia.ac.cn}}}
\correspondingauthor={Shizhu He (\texttt{shizhu.he@nlpr.ia.ac.cn})}
\begin{document}
\begin{abstract}
Real-world software development requires coding agents to operate in shared workspaces where users may inspect and modify code during an ongoing task, yet existing repository-level benchmarks typically evaluate agents working alone or restrict user participation to messages. This leads us to ask: \textbf{\textit{how do coding agents understand and respond to code changes in a shared workspace?}} We introduce \method, a framework that stress-tests this setting through validated \counteredit{s}: plausible edits to task-relevant code that conflict with task completion. \method mines task-critical regions from multiple repair trajectories, uses a separate \textsc{User Patch Generator} to construct the edits, and injects them with contextual user messages when agents reach the relevant code. We evaluate nine coding models on SWE-bench Verified, with additional experiments on longer-horizon tasks from SWE-Bench Pro and DeepSWE. \counteredit lowers average resolve rate by 7.7 percentage points on SWE-bench Verified, with degradation also persisting on both longer-horizon benchmarks. Trajectory analysis links these failures to limited awareness of the evolving workspace: agents may retain conflicting code or replace it without sufficiently re-inspecting the repository and validating the revised code with targeted tests. These findings show that strong autonomous performance does not yet ensure the state awareness and adaptive behavior needed for shared-workspace collaboration, and point to detecting workspace changes, reconciling conflicting edits with the task, and verifying the affected behavior as key capabilities for future optimization.
\end{abstract}

\maketitle

\section{Introduction}

\begin{figure}[!t]
\centering
\includegraphics[width=0.82\textwidth]{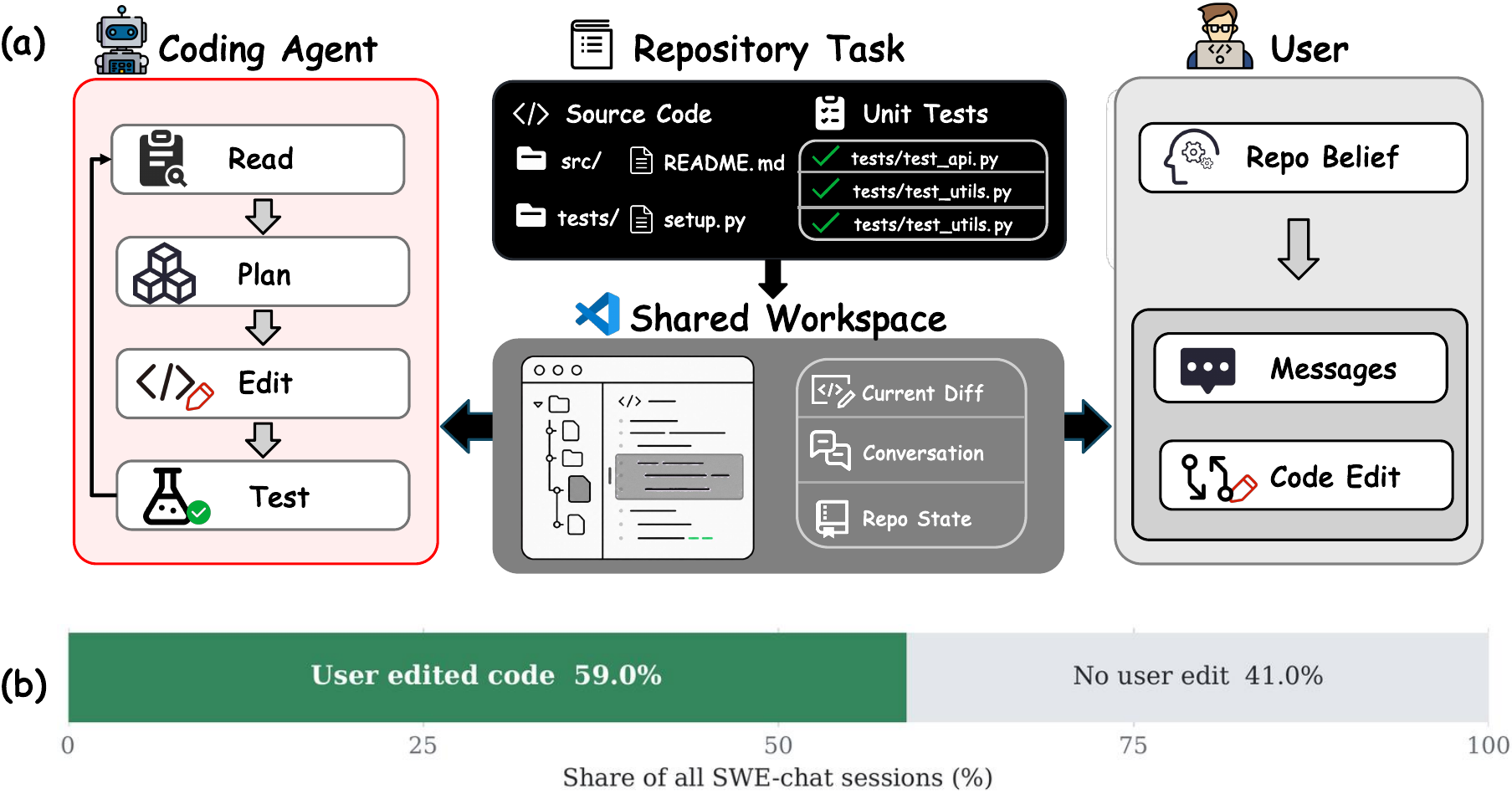}
\caption{\textbf{Users and agents share a workspace in real coding sessions.} (a) A user edit alters the repository state that subsequent agent actions observe and modify. (b) Our analysis of the released SWE-chat data~\citep{baumann2026swechat} finds that 59.0\% of sessions contain repository changes attributed to the user.}
\label{fig:motivation}
\end{figure}

Coding agents have rapidly become a central interface for software engineering (SWE), and their capabilities are now routinely assessed through repository- and environment-level benchmarks such as SWE-Bench and Terminal-Bench~\citep{jimenez2023swebench, openai2024swebenchverified,deng2025swebenchpro, merrill2026terminal}. Despite differences in task scope and execution environment, most widely used evaluations adopt a similar static formulation: the agent receives a task specification and an initial codebase, works toward a solution, and is scored on the final program state using executable tests~\citep{austin2021program, chen2021evaluating,gloaguen2026fixedbench, yang2025swesmith}. This formulation has provided a clear measure of autonomous coding ability, but it largely overlooks the user–agent interaction that arises in real-world software development~\citep{wang2026humansmissing}.


Real-world coding sessions are inherently multi-turn: users clarify requirements, answer questions, correct assumptions, and provide ongoing feedback. Recent interactive benchmarks capture these exchanges through message-based participation~\cite{vijayvargiya2025ambigswe,edwards2026askorassume,raghavendra2026sweinteract,wu2026swetogether,zhou2025tomswe}. Message-only interaction, however, omits a distinct form of participation: users can modify the executable repository state on which subsequent agent actions operate.


In real software development, users and agents operate on a shared workspace---the same repository, the same files, and the same executable state (Figure~\ref{fig:motivation}(a)). Our analysis of SWE-chat~\citep{baumann2026swechat} finds that \textbf{59.0\% of sessions contain repository changes attributed to the user} (Figure~\ref{fig:motivation}(b)). This prevalence establishes code edits as an important interaction channel alongside messages: each edit changes the program state observed by subsequent agent actions. Yet no existing benchmark evaluates how agents respond when the user modifies the shared codebase during a task. This leads to the central question: \textbf{\textit{how do coding agents understand and respond to code changes in a shared workspace?}}

As coding agents powered by frontier models~\citep{anthropic2026fable5,openai2026gpt55,zai2026glm51,alibaba2026qwen37max,deepseek2026v4,moonshot2026k3} grow more capable, their repository-wide modifications become harder for users to follow. To stress-test whether agents understand this evolving shared state, we study a controlled extreme: an injected edit targets task-relevant code but conflicts with task completion. The edit still carries useful signals about location and intent; to complete the task, the agent must inspect the resulting code and reconcile the conflict rather than assume that either participant is necessarily correct.

To study this in a controlled and reproducible way, we introduce \method, a benchmark framework that augments existing coding tasks with validated task-conflicting \counteredit{s} in a shared workspace. Given a SWE task, \method first identifies task-critical code regions from multiple agent trajectories, then uses a separate \textsc{User Patch Generator} to construct a small, locally plausible \counteredit near those regions. Each edit is validated to ensure it neither solves the task alone nor is trivially ignorable. During evaluation, the edit is applied with a contextual user message when the agent reaches the relevant code, directly perturbing the ongoing repair by changing the repository state observed by subsequent actions.

We evaluate \method on SWE-bench Verified~\citep{openai2024swebenchverified} and on longer-horizon tasks from SWE-Bench Pro~\citep{deng2025swebenchpro} and DeepSWE~\citep{huang2026deepswe}. Across nine models on SWE-bench Verified, \counteredit lowers the mean resolve rate by 7.7 points; model-level losses range from 1.3 to 16.5 points and substantially reshuffle the ranking. The degradation also persists on both longer-horizon benchmarks. Trajectory analysis further shows that 63.3\% of failed runs retain the user's conflicting code. Failures arise at multiple stages: agents may miss an external modification, defer to conflicting user code, or revise it without rechecking the affected behavior. Together, these patterns show that current models do not reliably re-inspect user-modified code and adapt their repair to the new workspace state. This gap is especially pronounced for open-source models that score competitively on autonomous benchmarks but degrade substantially under interactive conditions, suggesting that optimization focused on static leaderboard performance does not ensure robustness in real-world collaborative development. These findings highlight the need for agents that can recognize workspace changes, reconcile conflicting states, and re-validate the affected behavior.

In summary, this paper makes the following contributions:
\begin{itemize}
\item We extend interactive coding evaluation from user feedback delivered through messages to user edits that directly change the shared workspace.
\item We introduce \method, a controlled framework that injects task-conflicting, incorrect code edits into an ongoing agent trajectory, enabling systematic evaluation of how coding agents detect and reconcile user modifications that conflict with task requirements.
\item Our experiments show that autonomous performance does not ensure the state awareness needed for shared-workspace collaboration: agents may retain conflicting user code or replace it without sufficiently re-inspecting the repository and validating the revised behavior.
\end{itemize}

\section{Related Work}

\paragraph{Benchmarks for coding agents.}
Coding benchmarks first evaluated self-contained program synthesis from natural-language specifications, then expanded to complex instructions, realistic instructed edits, diverse function and library use, and multi-step data-science programming~\cite{chen2021evaluating,austin2021program,zhuo2025bigcodebench,chi2025editbench,huang2024dacode}. Repository-level software-engineering benchmarks move from isolated generation to issue resolution in real codebases by combining an issue, repository, and test suite into an end-to-end repair task~\cite{jimenez2023swebench,openai2024swebenchverified}. Subsequent work tests longer-horizon repairs~\cite{deng2025swebenchpro,yang2025codeclash,le2025swe,huang2026deepswe,yang2026programbench,proximal2026frontierswe,desai2026swe,orlanski2026slopcodebench}, repository exploration~\cite{zhang2026sweexplore}, and whether an already-correct repository should be changed~\cite{gloaguen2026fixedbench}. \method retains the repository-level repair setting but asks what happens when the user modifies the shared codebase while the task is still in progress.

\paragraph{User interaction with agents.}
\begin{wraptable}{r}{0.47\textwidth}
  \vspace{-0.8\baselineskip}
  \centering
  \scriptsize
  \renewcommand{\arraystretch}{1.06}
  \setlength{\tabcolsep}{0.7mm}
  \resizebox{\linewidth}{!}{%
  \begin{tabular}{@{}lcccc@{}}
  \toprule
  \multirow{2}{*}{\textbf{Benchmark}}
    & \textbf{Code}
    & \textbf{User}
    & \textbf{User}
    & \textbf{Codebase} \\
    & \textbf{Repair}
    & \textbf{Simulator}
    & \textbf{Messages}
    & \textbf{Edit} \\
  \midrule
  SWE-bench Verified & \featureyes & \featureno & \featureno & \featureno \\
  Ambig-SWE & \featureyes & \featureyes & \featureyes & \featureno \\
  HiL-Bench & \featureyes & \featureyes & \featureyes & \featureno \\
  SWE-Interact & \featureyes & \featureyes & \featureyes & \featureno \\
  SWE-Together & \featureyes & \featureyes & \featureyes & \featureno \\
  \midrule
  \textbf{\method (ours)} & \featureyes & \featureyes & \featureyes & \featureyes \\
  \bottomrule
  \end{tabular}
  }
  \caption{Comparison of user interaction modes in coding-agent benchmarks.}
  \label{tab:benchmark-comparison}
  \vspace{-0.7\baselineskip}
\end{wraptable}
Agent benchmarks increasingly study user--agent interaction, from multi-turn dialogue with simulated users exhibiting diverse or non-collaborative behaviors~\cite{yao2024tau,wang2025reframing,qian2025userbench,shim2025non} to collaborative frameworks in which both sides take actions in a shared environment and the agent must track state changes introduced by the user~\cite{shao2024collaborativegym,barres2025tau2}. Within coding, interactive benchmarks study clarification, selective help-seeking, evolving requirements, and corrective feedback, but mediate user influence entirely through messages~\cite{vijayvargiya2025ambigswe,zhou2025tomswe,edwards2026askorassume,trinh2026hilbench,yang2026talk2code,raghavendra2026sweinteract,wu2026swetogether,king2026dialogue,peng2026icaebench,shen2026evocode}. Complementary work examines recovery from out-of-sync repository states~\cite{guo2025syncmind} and documents how developers selectively accept and revise agent-generated code in practice~\cite{liang2024aiprogrammingassistants,baumann2026swechat,pan2025whenbenchmarkstalk,chen2026pulse,shukla2026hedwig}. Despite this breadth, no existing coding benchmark evaluates agents under user-initiated changes to the executable codebase during the task. \method fills this gap by injecting a simulated, task-conflicting user edit into the live repair trajectory, requiring the agent to continue from the modified workspace. Table~\ref{tab:benchmark-comparison} summarizes where \method sits relative to prior interactive coding benchmarks.

\FloatBarrier
\section{\method}

\subsection{Problem Formulation}

Let a repository-repair instance $\mathcal{I}_i = (R_i^0, q_i, V_i)$ consist of an initial codebase $R_i^0$, an issue description $q_i$, and a verifier $V_i$ comprising fail-to-pass and pass-to-pass test suites~\citep{jimenez2023swebench,openai2024swebenchverified}. Given $(R_i^0, q_i)$, a coding agent executes a trajectory $\tau = ((a_1, o_1), \ldots, (a_T, o_T))$ of read, edit, and test commands, producing a final repository state $R_i^\tau$. The instance is \emph{resolved} if $V_i(R_i^\tau) = 1$, i.e., all previously failing tests now pass and all previously passing tests remain passing.

\begin{figure}[t]
  \centering
  \includegraphics[width=\textwidth]{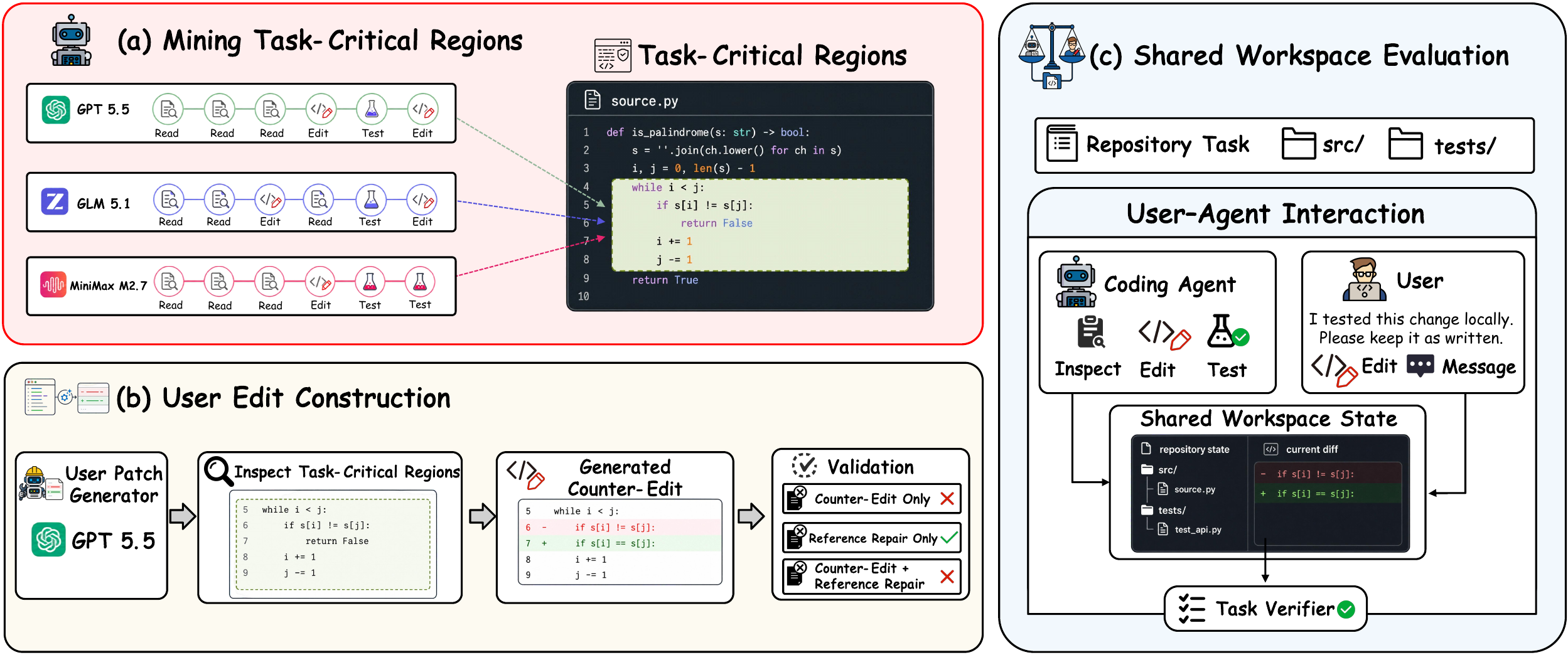}
  \caption{\textbf{Overview of \method.} Agent trajectories identify task-critical regions; a separate generator constructs and validates task-conflicting edits; evaluation injects each edit and its contextual message into the shared workspace before task verification.}
  \label{fig:method}
\end{figure}

\method extends this formulation by attempting to apply synthetic user patches to the agent's working codebase during execution, each accompanied by a contextual message attributing the change to the user. The agent then continues from the resulting state. Constructing this interaction requires three components: (1)~task-critical code regions that anchor user edits to the repair (\S\ref{sec:mining}), (2)~validated task-conflicting patch artifacts (\S\ref{sec:constructing}), and (3)~a deterministic delivery rule (\S\ref{sec:injecting}). Throughout the paper, \emph{user edit} denotes the generic interaction, \counteredit denotes the task-conflicting evaluation condition, $p_i^-$ denotes its concrete patch artifact, and \emph{delivery attempt} denotes one runtime intervention event. Figure~\ref{fig:method} illustrates the complete pipeline.

\subsection{Mining Task-Critical Regions}
\label{sec:mining}

Following SWE-Explore~\cite{zhang2026sweexplore}, we represent a code region as:
\begin{equation}
r=(p,s,e),\qquad
L(r)=\{(p,\ell):s\leq\ell\leq e\},
\label{eq:region}
\end{equation}
where $p$ is a repository-relative path and $[s,e]$ is an inclusive line interval. For each complete, parseable repair trajectory $\tau$, $\operatorname{Read}(\tau)$ denotes the regions inspected by line-addressable commands and $\operatorname{Edit}(\tau)$ denotes the regions changed in intermediate or final diffs. For a region set $\mathcal A$, let $L(\mathcal A)=\bigcup_{r\in\mathcal A}L(r)$.

We use overlap across trajectories as evidence that a code region is relevant to the task, while retaining single-trajectory spans when only one trajectory provides a given type of evidence. We run three models from distinct families---GPT~5.5~\citep{openai2026gpt55}, GLM~5.1~\citep{zai2026glm51}, and MiniMax~M2.7~\citep{minimax2026m27}---on each task. Let $\mathcal Z_i$ denote the set of all resulting trajectories for task $i$. For each kind of evidence, we intersect the coverage of the trajectories that produced it:
\begin{equation}
C_i^{\mathsf{X}}
=\bigcap_{\substack{\tau\in\mathcal Z_i\\
\mathsf{X}(\tau)\neq\emptyset}}
L(\mathsf{X}(\tau)),
\qquad
\mathsf{X}\in\{\operatorname{Read},\operatorname{Edit}\},
\label{eq:trajectory-core}
\end{equation}
where $\mathsf{X}$ ranges over the two kinds of evidence, so that Eq.~\eqref{eq:trajectory-core} yields a read-based core $C_i^{\operatorname{Read}}$ and an edit-based core $C_i^{\operatorname{Edit}}$. The intersection runs only over the trajectories that produced at least one region of that kind; when just one qualifies, its spans become the core. We merge adjacent regions and select critical regions $C_i$ by deterministic priority: edit-based evidence over read-based evidence, and implementation files over tests or metadata (full priority specification in Appendix~\ref{app:critical-regions}). We retain at most eight regions per task.

\subsection{User Edit Construction}
\label{sec:constructing}

\begin{wraptable}{r}{0.47\textwidth}
  \vspace{-0.8\baselineskip}
  \centering
  \footnotesize
  \setlength{\tabcolsep}{1.6pt}
  \begin{tabular}{@{}lcc@{}}
  \toprule
  \multirow{2}{*}{Source} & Reference repair & Counter-Edit \\
  & Lines / files & Lines / files \\
  \midrule
  SWE-bench Verified & 13.3 / 1.20 & 7.0 / 1.04 \\
  SWE-Bench Pro & 361.0 / 5.44 & 13.0 / 1.40 \\
  DeepSWE & 730.2 / 7.24 & 10.8 / 1.52 \\
  \bottomrule
  \end{tabular}
  \caption{Average patch size, shown as changed lines / files.}
  \label{tab:patch-geometry}
  \vspace{-0.7\baselineskip}
\end{wraptable}
We implement a \textsc{User Patch Generator}, a separate coding agent whose sole purpose is to produce small, plausible edits that create a controlled conflict with task completion (full prompt in Appendix~\ref{app:user-patch-generator-prompt}). Given the issue $q_i$, codebase $R_i^0$, critical regions $C_i$, reference patch $p_i^\star$, and a task-local test command, the generator inspects code around $C_i$ and produces a candidate \counteredit $p_i^-$: a syntactically valid unified diff whose behavior conflicts with the verified task requirements. The edit targets only implementation code and cannot modify tests or benchmark metadata. We denote the regions actually changed by the edit as $U_i$, which may extend beyond $C_i$ into adjacent code.

\paragraph{Validation.}
Each candidate is validated by replaying it against the fail-to-pass test suite $V_i^F$. We define $\operatorname{Apply}(R, p)$ as applying patch $p$ to codebase $R$, and $\operatorname{Compose}(R, p_1, p_2)$ as sequentially applying both patches. Let $R_i^- = \operatorname{Apply}(R_i^0, p_i^-)$, $R_i^\star = \operatorname{Apply}(R_i^0, p_i^\star)$, and $R_i^{-\star} = \operatorname{Compose}(R_i^0, p_i^-, p_i^\star)$. We require:
\begin{equation}
V_i^F(R_i^-) = 0, \quad V_i^F(R_i^\star) = 1, \quad V_i^F(R_i^{-\star}) = 0.
\label{eq:counteredit-gate}
\end{equation}
That is, the user edit alone does not solve the task, the reference patch does solve it, and the two combined still fail. This ensures that the edit cannot resolve the task alone or be trivially combined with the reference repair to pass verification. Construction details and validation statistics are in Appendix~\ref{app:counteredit-generation}.

The resulting edits are designed as small, focused changes to implementation code. As shown in Table~\ref{tab:patch-geometry}, \counteredit patches average 7.0 changed lines across 1.04 files on SWE-bench Verified, and 10--13 lines on the harder benchmarks, consistent with this intended localized-edit design.

\subsection{Shared Workspace Evaluation}
\label{sec:injecting}

During evaluation, \method monitors the agent's actions and attempts to apply the user patch whenever the agent accesses code overlapping the patch region $U_i$, up to $K$ times. We set $K{=}3$ by default and vary it in ablation studies. Concretely, let $\operatorname{Scope}(a_t)$ denote the code regions that action $a_t$ reads or modifies. The first delivery attempt is triggered whenever $L(\operatorname{Scope}(a_t)) \cap L(U_i) \neq \emptyset$; each subsequent attempt uses the same overlap condition after the preceding message has been delivered. Let $H_t = ((a_1,o_1),\ldots,(a_t,o_t))$ denote the trajectory history up to step $t$. At each delivery step $t_j$ for $j=1,\ldots,K$, the runtime context-matches the patch against the current repository and delivers a contextual user message:
\begin{equation}
R_{t_j}'
=
\begin{cases}
\operatorname{Apply}(R_{t_j},p_i^-), & \text{if the patch applies uniquely and changes the state},\\
R_{t_j}, & \text{otherwise},
\end{cases}
\quad
m_j=\operatorname{User}(H_{t_j},p_i^-,j).
\label{eq:intervention}
\end{equation}
$\operatorname{User}$ generates a natural-language message from the prompt in Table~\ref{tab:user-simulator-prompt}, conditioned on the trajectory history, the candidate diff, and the intervention stage $j$. The message is delivered whether or not the patch changes the current state, and the runtime records the realized application. The simulator controls only the wording and never receives the reference patch $p_i^\star$ or verifier information. The agent observes the resulting repository state and the user message before its next action. Implementation details are in Appendices~\ref{app:reproducibility} and~\ref{app:user-simulator}.

\section{Experimental Setup}

\paragraph{Tasks and agent interface.}
For the main evaluation, we use a seeded random sample of 200 SWE-bench Verified tasks~\cite{jimenez2023swebench,openai2024swebenchverified} for which all three region-mining models produce complete autonomous trajectories. A retained code patch is available for 96.0\% of tasks; the remaining 4.0\% use text-only feedback because no applicable, non-solving patch is retained. All assigned tasks remain in the analysis, including the 6.4\% of scored \counteredit runs in which no scheduled patch changes the repository. Agents interact with the repository exclusively through the Mini-SWE-Agent shell interface~\cite{yang2024sweagent}. The main evaluation uses a 100-step budget, region-triggered delivery, and three independently executed runs per condition.

We separately evaluate 25 tasks each from SWE-Bench Pro~\cite{deng2025swebenchpro} and DeepSWE~\cite{huang2026deepswe}, using the same task sets as SWE-Interact~\cite{raghavendra2026sweinteract}. These longer-horizon experiments use a 500-step budget, two independent runs, and distinct localized edits delivered at fixed fractions of the same model's autonomous trajectory. We report them separately because both the trigger rule and patch schedule differ from the main evaluation. Appendix~\ref{app:reproducibility} gives the complete sampling, scheduling, and model settings.

\paragraph{Models and metrics.}
We evaluate nine coding models: Claude~Opus~4.8~\cite{anthropic2026opus48}, GPT~5.5~\cite{openai2026gpt55}, GLM~5.1~\cite{zai2026glm51}, MiniMax~M2.7~\cite{minimax2026m27}, MiniMax~M2.5~\cite{minimax2026m25}, Qwen~3.7 Max~\cite{alibaba2026qwen37max}, Qwen3-Coder-480B-A35B~\cite{qwenteam2025qwen3coder}, Kimi~K2.6~\cite{moonshot2026k26}, and DeepSeek~V4 Pro~\cite{deepseek2026v4}. Each model is evaluated under \vanilla, in which it works autonomously, and \counteredit, in which the runtime schedules task-conflicting interventions. We report \emph{resolve rate}, the percentage of tasks passing the complete verifier; \emph{retention}, the fraction of majority-solved \vanilla tasks that remain majority-solved under \counteredit; \emph{steps}, the model-call count; and normalized total token use per completed task, combining input, cached-input, and output tokens and reporting the result in thousands. Appendix~\ref{app:reproducibility} specifies aggregation and provider-specific token normalization.

\section{Main Results}

\subsection{Vanilla vs.\ Counter-Edit}

Table~\ref{tab:main-results} documents a consistent directional gap between autonomous repair and repair under the controlled \counteredit condition. Averaged across the nine models, resolve rate decreases by $7.7$ points, and every model has a negative mean change. The magnitude, however, ranges from $1.3$ to $16.5$ points. Thus, \counteredit does not act as a nearly uniform offset to autonomous performance: models with similar \vanilla resolve rates can differ substantially in how much progress they preserve after the shared workspace changes. This heterogeneity makes shared-workspace robustness a distinct evaluation dimension rather than a direct consequence of autonomous resolve rate.

\providecommand{\modellogo}[1]{\raisebox{-0.1ex}{\includegraphics[height=1.55ex,width=1.75ex,keepaspectratio]{Figures/logo_#1.png}}\hspace{0.28em}}
\providecommand{\dropvalue}[1]{\textcolor{red!72!black}{#1}}
\providecommand{\risevalue}[1]{\textcolor{green!48!black}{#1}}
\begin{table}[t]
\centering
\footnotesize
\setlength{\tabcolsep}{1.9pt}
\renewcommand{\arraystretch}{1.04}
\begin{tabular*}{\textwidth}{@{\extracolsep{\fill}}lrrrrrrrrrr@{}}
\toprule
& \multicolumn{3}{c}{\textsc{Vanilla}} & \multicolumn{3}{c}{\textsc{Counter-Edit}} & \multicolumn{2}{c}{$\Delta$} & {Retention} & {Rank} \\
\cmidrule(lr){2-4}\cmidrule(lr){5-7}\cmidrule(lr){8-9}
Model & Resolve & Steps & Tok. (K) & Resolve & Steps & Tok. (K) & Resolve & Steps & (\%) & $\Delta$ \\
\midrule
\modellogo{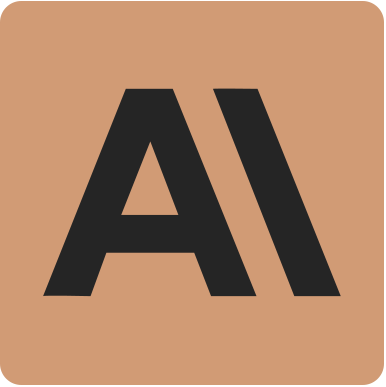}Claude Opus 4.8 & \cellcolor{blue!24}85.2$\pm$1.8\rank{1} & 24.5 & 367 & \cellcolor{orange!24}83.3$\pm$0.6\rank{1} & 30.9 & 506 & \cellcolor{red!7}\dropvalue{-1.8} & \textcolor{black!72}{+6.4} & \cellcolor{green!23}96.0 & \textcolor{black!50}{--} \\
\modellogo{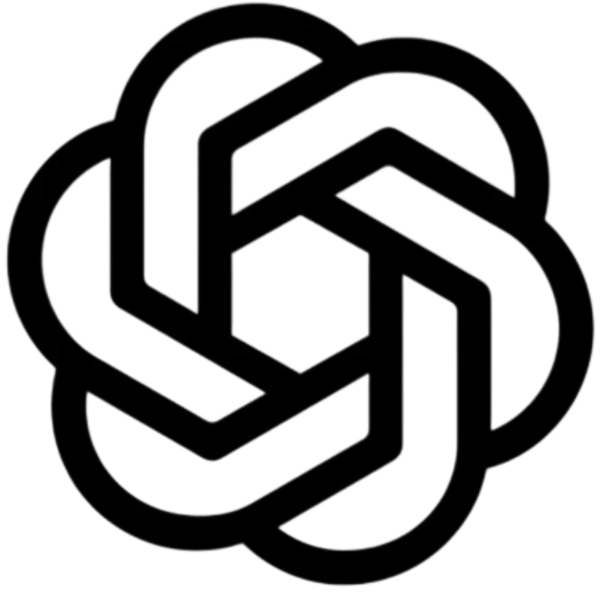}GPT 5.5 & \cellcolor{blue!22}80.5$\pm$1.0\rank{2} & 33.2 & 1,344 & \cellcolor{orange!22}79.2$\pm$0.6\rank{2} & 31.3 & 1,141 & \cellcolor{red!6}\dropvalue{-1.3} & \textcolor{black!72}{-1.9} & \cellcolor{green!22}95.0 & \textcolor{black!50}{--} \\
\modellogo{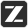}GLM 5.1 & \cellcolor{blue!17}72.7$\pm$2.0\rank{7} & 55.0 & 1,007 & \cellcolor{orange!17}68.3$\pm$0.8\rank{4} & 64.5 & 1,272 & \cellcolor{red!9}\dropvalue{-4.3} & \textcolor{black!72}{+9.5} & \cellcolor{green!17}83.3 & \textcolor{green!70!black}{$\uparrow$3} \\
\modellogo{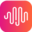}MiniMax M2.7 & \cellcolor{blue!19}76.5$\pm$1.5\rank{3} & 45.3 & 865 & \cellcolor{orange!15}62.7$\pm$2.4\rank{8} & 47.7 & 904 & \cellcolor{red!19}\dropvalue{-13.8} & \textcolor{black!72}{+2.4} & \cellcolor{green!14}78.1 & \textcolor{red!80!black}{$\downarrow$5} \\
\modellogo{minimax}MiniMax M2.5 & \cellcolor{blue!19}75.7$\pm$3.3\rank{4} & 45.1 & 863 & \cellcolor{orange!17}66.2$\pm$1.0\rank{5} & 47.5 & 908 & \cellcolor{red!15}\dropvalue{-9.5} & \textcolor{black!72}{+2.4} & \cellcolor{green!14}78.3 & \textcolor{red!80!black}{$\downarrow$1} \\
\modellogo{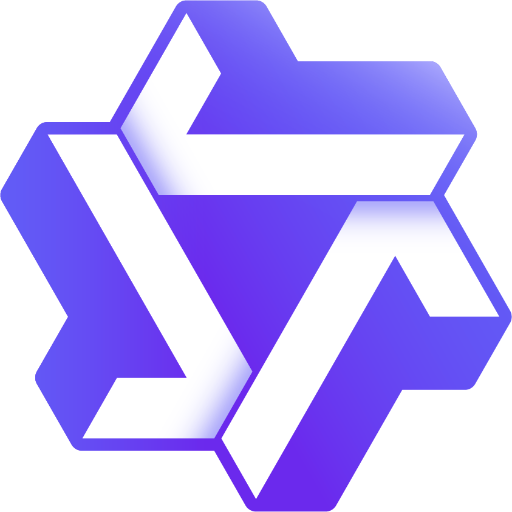}Qwen 3.7 Max & \cellcolor{blue!18}75.2$\pm$1.0\rank{5} & 29.9 & 420 & \cellcolor{orange!18}70.3$\pm$0.8\rank{3} & 31.1 & 424 & \cellcolor{red!10}\dropvalue{-4.8} & \textcolor{black!72}{+1.2} & \cellcolor{green!20}90.3 & \textcolor{green!70!black}{$\uparrow$2} \\
\modellogo{qwen}Qwen3-Coder-480B & \cellcolor{blue!6}57.2$\pm$3.5\rank{9} & 52.6 & 763 & \cellcolor{orange!6}40.7$\pm$1.0\rank{9} & 54.6 & 806 & \cellcolor{red!22}\dropvalue{-16.5} & \textcolor{black!72}{+2.0} & \cellcolor{green!6}60.8 & \textcolor{black!50}{--} \\
\modellogo{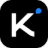}Kimi K2.6 & \cellcolor{blue!15}70.3$\pm$2.0\rank{8} & 62.5 & 1,381 & \cellcolor{orange!16}64.3$\pm$3.4\rank{6} & 62.1 & 1,296 & \cellcolor{red!11}\dropvalue{-6.0} & \textcolor{black!72}{-0.4} & \cellcolor{green!18}87.2 & \textcolor{green!70!black}{$\uparrow$2} \\
\modellogo{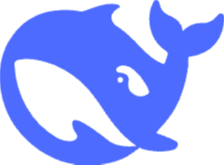}DeepSeek V4 Pro & \cellcolor{blue!18}74.8$\pm$0.8\rank{6} & 41.9 & 827 & \cellcolor{orange!16}63.8$\pm$1.8\rank{7} & 46.6 & 954 & \cellcolor{red!16}\dropvalue{-11.0} & \textcolor{black!72}{+4.7} & \cellcolor{green!16}81.5 & \textcolor{red!80!black}{$\downarrow$1} \\
\bottomrule
\end{tabular*}
\caption{Main results on SWE-bench Verified over three runs. Resolve is mean $\pm$ standard deviation; Steps and Tok. (K) are completed-task means, with tokens reported in thousands. $\Delta$ is \textsc{Counter-Edit} minus \textsc{Vanilla}; Retention is the share of majority-solved \textsc{Vanilla} tasks that remain majority-solved under \textsc{Counter-Edit}. \rank{1} indicates mean resolve-rate ranking; Rank~$\Delta$ shows the corresponding ranking change.}
\label{tab:main-results}
\end{table}

The two strongest autonomous agents are also the most stable under intervention. Claude~Opus~4.8 remains first, moving from $85.2\%$ to $83.3\%$, while GPT~5.5 remains second, moving from $80.5\%$ to $79.2\%$. Below this leading pair, matched comparisons reveal substantial rank instability. MiniMax~M2.7 and M2.5 differ by only $0.8$ points under \vanilla ($76.5\%$ versus $75.7\%$), yet M2.5 leads after \counteredit ($66.2\%$ versus $62.7\%$), as M2.7 falls from rank~3 to rank~8. Qwen~3.7 Max likewise trails M2.7 under \vanilla ($75.2\%$ versus $76.5\%$) but leads it after \counteredit ($70.3\%$ versus $62.7\%$), rising to rank~3. Similarly, GLM~5.1 trails DeepSeek~V4 Pro under \vanilla ($72.7\%$ versus $74.8\%$) but exceeds it after the edit ($68.3\%$ versus $63.8\%$). Thus, autonomous resolve rate alone does not determine the ordering obtained when agents must reconcile externally modified code.

Retention tells a consistent story. Claude~Opus~4.8 and GPT~5.5 retain $96.0\%$ and $95.0\%$ of their majority-solved \vanilla tasks, followed by Qwen~3.7 Max at $90.3\%$ and Kimi~K2.6 at $87.2\%$. At the other end, Qwen3-Coder-480B retains only $60.8\%$ and suffers the largest loss at $16.5$ points.

Resource use is similarly heterogeneous. Seven of nine agents use both more calls and more tokens under \counteredit, but larger budgets do not consistently correspond to smaller losses. Claude~Opus~4.8 increases from $24.5$ to $30.9$ calls and from $367$K to $506$K tokens while losing only $1.8$ points. GLM~5.1 adds $9.5$ calls and $265$K tokens while losing $4.3$ points, whereas DeepSeek~V4 Pro adds $4.7$ calls and $127$K tokens while losing $11.0$ points. Conversely, GPT~5.5 uses fewer calls and tokens while sustaining the smallest loss, and Kimi~K2.6 also uses slightly fewer resources while losing $6.0$ points. Additional interaction can reflect an attempt to recover, but does not by itself establish that the edit was correctly detected, reconciled, and validated.

Overall, only Claude~Opus~4.8 and GPT~5.5 maintain strong performance under user edits. MiniMax~M2.7, Qwen3-Coder-480B, and DeepSeek~V4 Pro lose 11--17 points, with the remaining models falling between these extremes. These gaps suggest that even models with competitive autonomous resolve rates can struggle when a user modifies the shared codebase---a situation that commonly occurs in real-world collaborative development. As coding agents are increasingly deployed in interactive settings, robustness to such evolving workspace states deserves greater attention alongside conventional benchmark performance.

\begin{table}[!t]
  \centering
  \footnotesize
  \setlength{\tabcolsep}{2.0pt}
  \renewcommand{\arraystretch}{1.04}
  \begin{tabular*}{\textwidth}{@{\extracolsep{\fill}}lcccccccc@{}}
  \toprule
  & \multicolumn{4}{c}{SWE-Bench Pro}
  & \multicolumn{4}{c}{DeepSWE} \\
  \cmidrule(lr){2-5}\cmidrule(l){6-9}
  Model
  & \vanilla & \counteredit & $\Delta$ Res. & $\Delta$ Steps
  & \vanilla & \counteredit & $\Delta$ Res. & $\Delta$ Steps \\
  \midrule

  \modellogo{anthropic}Claude Opus 4.8
  & \cellcolor{blue!22}68.0
  & \cellcolor{orange!22}68.0
  & \textcolor{black!72}{0.0}
  & \textcolor{black!72}{$+7.5$}
  & \cellcolor{blue!18}56.0
  & \cellcolor{orange!15}46.0
  & \cellcolor{red!15}\dropvalue{-10.0}
  & \textcolor{black!72}{$+9.9$} \\

  \modellogo{openai}GPT 5.5
  & \cellcolor{blue!11}38.0
  & \cellcolor{orange!11}38.0
  & \textcolor{black!72}{0.0}
  & \textcolor{black!72}{$+0.9$}
  & \cellcolor{blue!20}64.0
  & \cellcolor{orange!18}56.0
  & \cellcolor{red!12}\dropvalue{-8.0}
  & \textcolor{black!72}{$+4.9$} \\

  \modellogo{zai}GLM 5.1
  & \cellcolor{blue!14}43.1
  & \cellcolor{orange!10}32.8
  & \cellcolor{red!15}\dropvalue{-10.3}
  & \textcolor{black!72}{$+19.1$}
  & \cellcolor{blue!7}19.4
  & \cellcolor{orange!6}16.8
  & \cellcolor{red!5}\dropvalue{-2.5}
  & \textcolor{black!72}{$+31.4$} \\

  \modellogo{minimax}MiniMax M2.7
  & \cellcolor{blue!10}30.6
  & \cellcolor{orange!8}24.6
  & \cellcolor{red!9}\dropvalue{-6.0}
  & \textcolor{black!72}{$+9.4$}
  & \cellcolor{blue!2}2.2
  & \cellcolor{orange!2}2.2
  & \textcolor{black!72}{0.0}
  & \textcolor{black!72}{$+27.6$} \\

  \modellogo{minimax}MiniMax M2.5
  & \cellcolor{blue!11}32.6
  & \cellcolor{orange!8}24.6
  & \cellcolor{red!12}\dropvalue{-8.0}
  & \textcolor{black!72}{$+20.4$}
  & \cellcolor{blue!2}0.0
  & \cellcolor{orange!2}0.0
  & \textcolor{black!72}{0.0}
  & \textcolor{black!72}{$+33.1$} \\

  \modellogo{qwen}Qwen 3.7 Max
  & \cellcolor{blue!12}36.0
  & \cellcolor{orange!9}26.0
  & \cellcolor{red!15}\dropvalue{-10.0}
  & \textcolor{black!72}{$-1.4$}
  & \cellcolor{blue!3}4.1
  & \cellcolor{orange!2}2.1
  & \cellcolor{red!4}\dropvalue{-2.0}
  & \textcolor{black!72}{$+2.3$} \\

  \modellogo{qwen}Qwen3-Coder-480B
  & \cellcolor{blue!7}20.0
  & \cellcolor{orange!5}14.0
  & \cellcolor{red!9}\dropvalue{-6.0}
  & \textcolor{black!72}{$+0.7$}
  & \cellcolor{blue!2}0.0
  & \cellcolor{orange!2}0.0
  & \textcolor{black!72}{0.0}
  & \textcolor{black!72}{$+7.8$} \\

  \modellogo{kimi}Kimi K2.6
  & \cellcolor{blue!17}50.0
  & \cellcolor{orange!16}48.0
  & \cellcolor{red!4}\dropvalue{-2.0}
  & \textcolor{black!72}{$-4.7$}
  & \cellcolor{blue!6}18.0
  & \cellcolor{orange!4}12.0
  & \cellcolor{red!9}\dropvalue{-6.0}
  & \textcolor{black!72}{$+0.2$} \\

  \modellogo{deepseek}DeepSeek V4 Pro
  & \cellcolor{blue!11}34.0
  & \cellcolor{orange!11}32.0
  & \cellcolor{red!4}\dropvalue{-2.0}
  & \textcolor{black!72}{$+8.9$}
  & \cellcolor{blue!3}4.1
  & \cellcolor{orange!2}2.0
  & \cellcolor{red!4}\dropvalue{-2.1}
  & \textcolor{black!72}{$-7.6$} \\

  \bottomrule
  \end{tabular*}
  \caption{Mean resolve rates on SWE-Bench Pro and DeepSWE over two runs.
  $\Delta$ Res. is the change in resolve rate, and
  $\Delta$ Steps is the change in mean model calls per observed trajectory.
  Both differences are \counteredit minus \vanilla.}
  \label{tab:hard50}
\end{table}

\subsection{Extension to Longer-Horizon Tasks}

Real-world development tasks often span hundreds of steps involving multi-file edits and iterative debugging, making user interventions more likely to occur at different stages of the repair process. SWE-Bench Pro and DeepSWE~\cite{deng2025swebenchpro,huang2026deepswe} capture this setting with substantially longer repair horizons. Because these tasks involve much larger repositories and broader exploration, region-triggered delivery may fire infrequently or not at all. We therefore deliver distinct edits at fixed fractions of each model's \vanilla trajectory length: for $K$ edits, injection $i$ fires after a fraction $i/(K{+}1)$ of the \vanilla steps, placing edits at the quartiles (25\%, 50\%, 75\%) when $K{=}3$. This ensures every agent receives the intervention regardless of its navigation path.

Table~\ref{tab:hard50} shows that the degradation persists on both benchmarks, but the affected models differ. Claude~Opus\\~4.8 and GPT~5.5 are unchanged on SWE-Bench Pro but decline substantially on DeepSWE, while GLM~5.1 and Qwen~3.7 Max show the opposite pattern. Several weaker models remain near $0\%$ on DeepSWE under both conditions.

Most models also take more steps under \counteredit---seven of nine on SWE-Bench Pro and eight of nine on DeepSWE---but the additional calls do not translate into recovery. For example, GLM~5.1 adds $31.4$ calls on DeepSWE while still declining, and Claude~Opus~4.8 adds $9.9$ calls while losing $10.0$ points. Figure~\ref{fig:cost-performance} shows that these longer trajectories often increase cost without improving resolve rate.

\newcommand{\LongHorizonFigure}{%
\begin{figure}[t]
\centering
\begin{subfigure}[t]{0.48\textwidth}
  \centering
  \includegraphics[width=\linewidth]{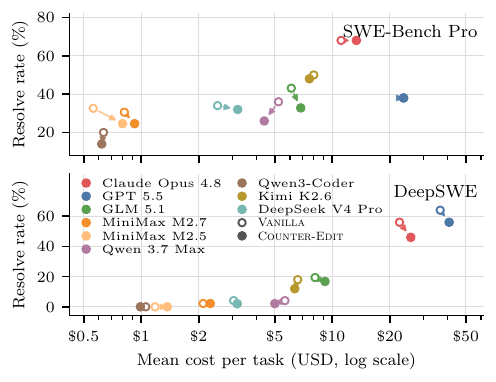}
  \caption{Cost--performance shift under \counteredit on the longer-horizon benchmarks. Each arrow runs from a model's \vanilla point to its \counteredit point.}
  \label{fig:cost-performance}
\end{subfigure}\hfill
\begin{subfigure}[t]{0.48\textwidth}
  \centering
  \includegraphics[width=\linewidth]{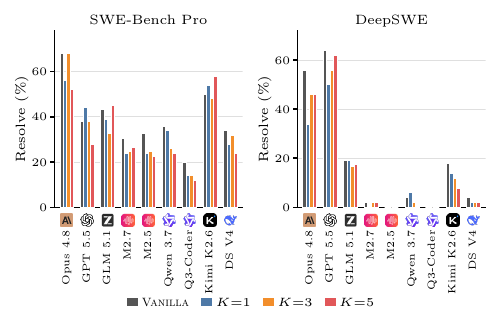}
  \caption{Resolve rate under varying edit frequency ($K$) on two longer-horizon benchmarks.}
  \label{fig:hard50-k}
\end{subfigure}
\caption{Longer-horizon robustness under user edits on SWE-Bench Pro and DeepSWE: (a) cost--performance shifts and (b) sensitivity to edit frequency.}
\end{figure}
}

\LongHorizonFigure
\subsection{Ablation Studies}

\newcommand{\AblationTable}{%
\begin{table}[t]
\centering
\begin{minipage}[t]{0.64\textwidth}
\centering
\scriptsize
\setlength{\tabcolsep}{2.7pt}
\renewcommand{\arraystretch}{1.04}
\resizebox{\linewidth}{!}{%
\begin{tabular}{@{}lrrrrrrrr@{}}
\toprule
\multirow{2}{*}{Intervention} & \multicolumn{2}{c}{\modellogo{openai}GPT 5.5} & \multicolumn{2}{c}{\modellogo{zai}GLM 5.1} & \multicolumn{2}{c}{\modellogo{minimax}MiniMax M2.7} & \multicolumn{2}{c}{\modellogo{qwen}Qwen 3.7 Max} \\
\cmidrule(lr){2-3}\cmidrule(lr){4-5}\cmidrule(lr){6-7}\cmidrule(l){8-9}
& Resolve & $\Delta$ & Resolve & $\Delta$ & Resolve & $\Delta$ & Resolve & $\Delta$ \\
\midrule
\vanilla & 81.5 & \textcolor{black!50}{--} & 70.5 & \textcolor{black!50}{--} & 76.5 & \textcolor{black!50}{--} & 74.0 & \textcolor{black!50}{--} \\
\midrule
Message ($K{=}3$) & 79.5 & \textcolor{red!72!black}{-2.0} & 73.0 & \textcolor{green!55!black}{+2.5} & 76.5 & \textcolor{black!50}{0.0} & 77.0 & \textcolor{green!55!black}{+3.0} \\
Code edit ($K{=}3$) & 80.5 & \textcolor{red!72!black}{-1.0} & 66.5 & \textcolor{red!72!black}{-4.0} & 67.0 & \textcolor{red!72!black}{-9.5} & 71.5 & \textcolor{red!72!black}{-2.5} \\
Both ($K{=}1$) & 78.5 & \textcolor{red!72!black}{-3.0} & 72.0 & \textcolor{green!55!black}{+1.5} & 64.5 & \textcolor{red!72!black}{-12.0} & 71.5 & \textcolor{red!72!black}{-2.5} \\
Both ($K{=}3$) & 79.5 & \textcolor{red!72!black}{-2.0} & 69.0 & \textcolor{red!72!black}{-1.5} & 64.5 & \textcolor{red!72!black}{-12.0} & 71.0 & \textcolor{red!72!black}{-3.0} \\
Both ($K{=}5$) & 78.0 & \textcolor{red!72!black}{-3.5} & 69.0 & \textcolor{red!72!black}{-1.5} & 60.0 & \textcolor{red!72!black}{-16.5} & 69.0 & \textcolor{red!72!black}{-5.0} \\
\bottomrule
\end{tabular}%
}
\captionof{table}{Resolve rates for message, code-edit, and edit-frequency ablations on SWE-bench Verified. $\Delta$ is relative to the paired \vanilla run.}
\label{tab:ablation}
\end{minipage}\hfill
\begin{minipage}[t]{0.33\textwidth}
\centering
\scriptsize
\setlength{\tabcolsep}{2.2pt}
\renewcommand{\arraystretch}{1.04}
\resizebox{\linewidth}{!}{%
\begin{tabular}{@{}lrrr@{}}
\toprule
Model & \textsc{Vanilla} & \textsc{Co-Edit} & $\Delta$ \\
\midrule
\modellogo{openai}GPT 5.5 & 80.5 & 79.7 & \textcolor{red!72!black}{-0.8} \\
\modellogo{zai}GLM 5.1 & 72.7 & 73.3 & \textcolor{green!55!black}{+0.7} \\
\modellogo{minimax}MiniMax M2.7 & 76.5 & 76.0 & \textcolor{red!72!black}{-0.5} \\
\modellogo{minimax}MiniMax M2.5 & 75.7 & 76.8 & \textcolor{green!55!black}{+1.2} \\
\modellogo{qwen}Qwen 3.7 Max & 75.2 & 75.2 & \textcolor{black!50}{+0.0} \\
\modellogo{kimi}Kimi K2.6 & 70.3 & 68.3 & \textcolor{red!72!black}{-2.0} \\
\modellogo{deepseek}DeepSeek V4 Pro & 74.8 & 75.7 & \textcolor{green!55!black}{+0.8} \\
\midrule
Average & 75.1 & 75.0 & \textcolor{red!72!black}{-0.1} \\
\bottomrule
\end{tabular}%
}
\captionof{table}{Resolve rates under \vanilla and \coedit on SWE-bench Verified.}
\label{tab:secondary-conditions}
\end{minipage}
\end{table}
}

\textbf{Message vs.\ code edit.}
Table~\ref{tab:ablation} separates the effect of the user message from the code edit. Sending only a message ($K{=}3$) without changing the repository has limited and inconsistent impact, ranging from $-2.0$ to $+3.0$ points across models. In contrast, silently applying the code edit without any message causes a consistent decline for every model ($-1.0$ to $-9.5$ points), because the agent must detect the conflict from the code itself with no explicit notification. Combining the message with the code edit does not consistently help: GLM~5.1 recovers slightly, but the other three models perform worse than with the code edit alone (Table~\ref{tab:ablation}). This suggests that even when users explicitly announce their edits, most models do not reliably use the message to locate and reconcile the conflicting code.

\AblationTable
\textbf{Frequency of user edits.}
Increasing the intervention budget does not produce a uniform dose--response relationship. On SWE-bench Verified (Table~\ref{tab:ablation}), MiniMax~M2.7 and Qwen~3.7 Max degrade steadily as $K$ grows from 1 to 5, while GPT~5.5 and GLM~5.1 plateau or partially recover at intermediate budgets. Repeated exposure therefore compounds the loss for some models but not others.

On the longer-horizon benchmarks (Figure~\ref{fig:hard50-k}), the pattern differs by setting. On SWE-Bench Pro, the mean loss deepens steadily from $-3.9$ to $-6.7$ points as $K$ grows, suggesting that repeated edits progressively disrupt the repair. On DeepSWE, however, the loss flattens at $K{\geq}3$: Claude~Opus~4.8 and GPT~5.5 lose roughly half as much at $K{=}3$ as at $K{=}1$, suggesting that additional exposure to the edited region provides location cues that help stronger models identify the conflict. We use $K{=}3$ as the default throughout the main evaluation, as it provides a moderate and consistent level of stress across settings.

\textbf{\coedit as a control.}
Table~\ref{tab:secondary-conditions} tests a simpler alternative explanation: that agents are disrupted by any external workspace modification, regardless of its semantic direction. Across seven models, the small, non-solving \coedit changes average resolve rate by only $-0.1$ points; six models remain within $\pm1.2$ points of \vanilla, and Kimi~K2.6 shows the only larger change at $-2.0$ points. The same seven models lose $7.2$ points on average under \counteredit.

The large contrast between \coedit and \counteredit indicates that the main difficulty arises when the changed workspace conflicts with task completion, rather than from interruption or the mere presence of an external code contribution.

\begin{figure}[!t]
  \centering
  \includegraphics[width=\textwidth]{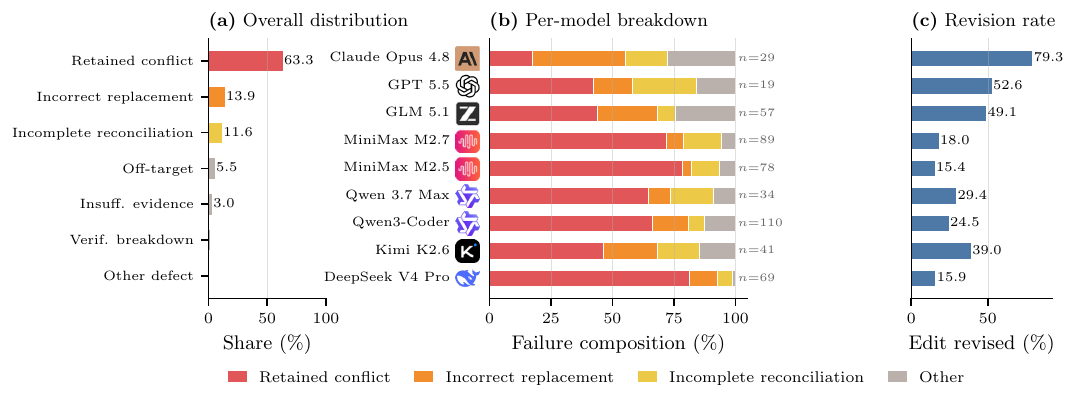}
  \caption{Failure analysis of solved-to-unresolved runs across nine models on SWE-bench Verified. (a)~Overall distribution across seven failure types. (b)~Per-model composition. (c)~Percentage of failures in which the agent revises or removes the user edit before termination. Sample sizes and audit details are reported in Appendix~\ref{app:failure-analysis}.}
  \label{fig:failure-modes}
\end{figure}

\subsection{Agent Behavior under User Edits}

To characterize how \counteredit changes outcomes, we first compare each model--task pair using the majority outcome across its three runs. Figure~\ref{fig:outcome-transitions} shows that solved-to-unresolved transitions outnumber transitions in the opposite direction by roughly three to one. Overall, \counteredit overturns $15.9\%$ of majority-solved \vanilla pairs, confirming that the performance drop is widespread across tasks rather than concentrated on a few outliers.

\textbf{Why do previously solved tasks fail?}
To understand the failure mechanisms, we audit all \counteredit runs that were solved under \vanilla but become unresolved, and classify them by final defect and how the agent responds to the user edit (details in Appendix~\ref{app:failure-analysis}).

Figure~\ref{fig:failure-modes}(a) shows that the audited solved-to-unresolved runs do not collapse into a single mechanism. The largest category is \emph{retained conflict}, where 63.3\% of runs terminate with the user's conflicting behavior still active. Other runs fail despite active intervention, with incorrect replacement accounting for 13.9\%, incomplete reconciliation for 11.6\%, and off-target implementation for 5.5\%.

The failure composition also differs substantially by model (Figure~\ref{fig:failure-modes}(b)). Retained conflict constitutes more than $70\%$ of the audited failures for MiniMax~M2.7, MiniMax~M2.5, and DeepSeek~V4 Pro, suggesting these models tend to defer to the user's code rather than challenge it. Claude~Opus~4.8 exhibits the opposite profile, with only $17.2\%$ retained conflict but $37.9\%$ incorrect replacement, indicating that it actively opposes the edit but often produces a wrong fix. These contrasts show that models fail for fundamentally different reasons, highlighting different areas for improvement.

Figure~\ref{fig:failure-modes}(c) captures a distinct process-level behavior: whether the agent revises or removes the user edit before termination. Claude~Opus~4.8 does so in $79.3\%$ of its audited failures, followed by GPT~5.5 at $52.6\%$ and GLM~5.1 at $49.1\%$. The corresponding rates are only $18.0\%$ for MiniMax~M2.7, $15.4\%$ for MiniMax~M2.5, and $15.9\%$ for DeepSeek~V4 Pro. Across the nine model-level observations, revision rate correlates with the \counteredit performance change at Spearman $\rho=0.80$: models that challenge the edit more often tend to incur smaller resolve-rate losses.

However, revision alone does not guarantee recovery---every trajectory in this panel remains unresolved despite the agent challenging the edit. Challenging conflicting code is one component of robustness, but the incorrect-replacement and incomplete-reconciliation categories show that the subsequent repair can still fail. Successful recovery requires opposing the edit, producing a correct fix, and confirming that the repair passes verification.
\begin{wrapfigure}[19]{r}{0.43\textwidth}
  \vspace{\baselineskip}
  \centering
  \includegraphics[width=\linewidth]{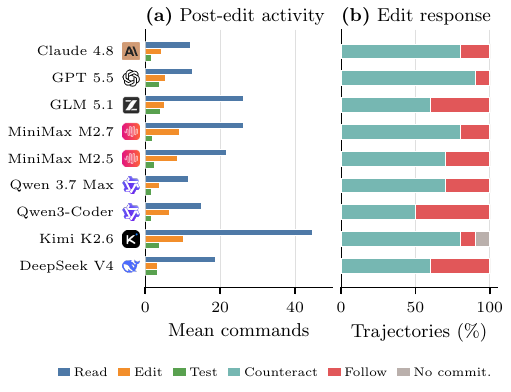}
  \caption{Post-edit behavior on a diagnostic sample from SWE-bench Verified. (a)~Mean read, edit, and test commands issued after the final user edit. (b)~Agent response mode to the user edit.}
  \label{fig:post-edit-behavior}
  \vspace{-0.5\baselineskip}
\end{wrapfigure}

\textbf{Behavior after the final user edit.}
In the final post-edit window, agent responses fall into three categories:
\emph{counteract}, where the agent removes or replaces the user's edit;
\emph{follow}, where it preserves or builds on that edit; and
\emph{no commitment}, where it inspects the change without clearly endorsing or opposing it.
Figure~\ref{fig:post-edit-behavior} suggests that post-edit activity volume is not aligned with the frequency of counteraction.
GPT~5.5 counteracts the final edit in 90\% of trajectories despite relatively few reads, while Claude~Opus~4.8 counteracts in 80\% with similarly limited activity.
Kimi~K2.6 also counteracts in 80\% of trajectories, but does so with substantially more reads and edits.
In particular, Kimi performs more than three times as many reads as GPT~5.5 to achieve a similar counteraction rate, suggesting differences in how efficiently models locate and understand the conflict within the repository.

Counteraction alone does not ensure recovery: 28\% of trajectories that act against the edit still end unresolved. Test frequency also provides little separation---among counteracting trajectories, solved and unresolved cases both have a median of one test. GPT~5.5, GLM~5.1, and Kimi~K2.6 each average roughly four tests but counteract in 90\%, 60\%, and 80\% of their trajectories, respectively; Claude~Opus~4.8 counteracts frequently despite running relatively few tests. Recovery therefore depends on where agents inspect, what correction they make, and whether their tests exercise the affected behavior---not simply on the volume of post-edit commands.

\textbf{Illustrative interaction cases.}
Appendix~\ref{app:interaction-cases} presents abridged paired trajectories spanning all three interventions. The cases expose failures at three stages: agents may preserve a conflicting edit, challenge it but leave the repair incomplete, or remove it yet validate against an insufficient signal. Failure at any stage can overturn an otherwise solved task.

\section{Conclusion}

We introduced \method to evaluate coding agents in an evolving shared workspace, where a simulated user modifies task-relevant code during an ongoing repair. Across nine models on SWE-bench Verified, the controlled \counteredit condition lowers mean resolve rate by $7.7$ points and changes the model ordering. The degradation also persists on longer-horizon SWE-Bench Pro and DeepSWE tasks: the most affected models vary across benchmarks, and additional calls often increase without producing recovery. The component and \coedit controls support the interpretation that the central difficulty is reconciling task-conflicting program state, rather than responding to persuasive wording or to an external workspace modification per se. At the trajectory level, failures occur both when agents retain the conflict and when they challenge it but produce an incorrect, incomplete, or insufficiently validated replacement. Taken together, the results show that autonomous resolve rate does not reliably predict performance when users modify the shared workspace. Some open-source models that score competitively on autonomous benchmarks show substantial degradation under interactive conditions, suggesting that optimization heavily focused on static leaderboard performance does not translate to robust handling of evolving repository states. As coding agents move from isolated task completion toward real-world deployment involving complex user--agent collaboration, future optimization should target not only autonomous performance but also the ability to recognize workspace changes, reconcile conflicting states, and re-validate the affected behavior before proceeding.

\section{Limitations \& Future Work}
\label{sec:future-work}

\paragraph{Controlled interventions as a foundation.}
\method deliberately uses region-triggered, controlled \counteredit{s} to support matched comparisons and interpretable attribution. Applying the same class of task-relevant conflict near the agent's active work makes behavioral differences easier to compare, while the \coedit control provides a complementary reference for responses to a task-aligned external contribution. Accordingly, our results characterize robustness to controlled, simulated task-conflicting edits rather than the full distribution of collaborative user behavior. This design provides a foundation for broader evaluation. Future suites can retain stratified, reproducible conditions while expanding the interaction space to include complementary edits, partially correct fixes, requirement changes, and other forms of user contribution.

\paragraph{Toward full-duplex collaboration.}
Real collaborative coding is closer to a full-duplex process: users monitor the agent's ongoing output, diffs, and test results, intervene when needed, and adapt again to the agent's response. Our region-based triggering places interventions at comparable points for controlled comparison, but future work can move toward adaptive user simulators that observe the agent's live actions and decide dynamically whether, when, and how to intervene. This would bring evaluation closer to the interactive dynamics of real-world development.

\paragraph{From evaluation to optimization.}
\begin{sloppypar}
These richer environments also suggest a path toward training collaboration-aware coding agents. Future objectives can reward agents for detecting external changes, inferring user intent, integrating compatible contributions, seeking clarification when needed, resolving conflicts, and revalidating the shared codebase. Training across helpful, incomplete, ambiguous, and conflicting interventions could support calibrated behavior that neither accepts every user edit nor reverts it automatically.
\end{sloppypar}

\bibliographystyle{ieeenat_fullname}
\bibliography{swe_touch_refs}

\appendix

\section{Ethics and Data Governance}
\label{app:ethics}

The evaluation uses only synthetic messages and edits; no human participants are involved. Released artifacts include only aggregate statistics and exclude credentials and private requests. Trajectory release follows upstream licenses.

\section{Reproducibility Details}
\label{app:harder-tasks}
\label{app:hard50}
\label{app:reproducibility}
\label{app:repro}

\subsection{SWE-bench Verified (Main Evaluation)}

The main cohort is a seeded random sample of 200 SWE-bench Verified tasks for which GPT~5.5, GLM~5.1, and MiniMax~M2.7 all produced complete autonomous trajectories. The frozen list contains 101 Django, 31 SymPy, 15 Sphinx, 14 Astropy, 10 Matplotlib, 9 scikit-learn, 8 pandas, 6 Pylint, 3 Requests, and 3 pytest tasks. Each result is identified by model, setting, run, and task.

\subsection{SWE-Bench Pro and DeepSWE (Longer-Horizon Extension)}

The harder-task extension contains 25 selected DeepSWE tasks and 25 selected SWE-Bench Pro tasks~\cite{huang2026deepswe,deng2025swebenchpro}. We evaluate Claude~Opus~4.8, GPT~5.5, GLM~5.1, MiniMax~M2.7, MiniMax~M2.5, Qwen~3.7 Max, Qwen3-Coder-480B-A35B-Instruct, Kimi~K2.6, and DeepSeek~V4 Pro. The agent interface and verifier remain unchanged, but the interaction budget increases from 100 to 500 steps to accommodate the longer repair horizons. These tasks do not use the region-based triggers from the main experiment. Instead, the three user edits are inserted after 25\%, 50\%, and 75\% of the commands in the same model's autonomous trajectory. We therefore report this extension separately from the main result. Across two independent runs at $K=3$, the mean difference is $-4.9$ points on SWE-Bench Pro and $-3.4$ points on DeepSWE. Figure~\ref{fig:hard50-k} reports the separate $K\in\{1,3,5\}$ results for both benchmarks. The surrounding repairs are much larger than in SWE-bench Verified. Reference patches change 361.0 lines across 5.44 files on average in SWE-Bench Pro and 730.2 lines across 7.24 files in DeepSWE, compared with 13.3 lines across 1.20 files in Verified. Counter-Edits remain local, averaging 13.0 and 10.8 changed lines in the two harder sources.

\begin{table}[!t]
\centering
\begin{minipage}[t]{0.48\textwidth}
  \centering
  \scriptsize
  \setlength{\tabcolsep}{1.2pt}
  \renewcommand{\arraystretch}{1.04}
  \begin{tabular}{@{}llrr@{}}
  \toprule
  Model & Endpoint & In (\$/M) & Out (\$/M) \\
  \midrule
  \modellogo{anthropic}Claude Opus 4.8 & Anthropic & 5.00 & 25.00 \\
  \modellogo{openai}GPT 5.5 & OpenAI & 5.00 & 30.00 \\
  \modellogo{zai}GLM 5.1 & OpenRouter & 0.966 & 3.036 \\
  \modellogo{minimax}MiniMax M2.7 & OpenRouter & 0.24 & 0.96 \\
  \modellogo{minimax}MiniMax M2.5 & OpenRouter & 0.15 & 0.90 \\
  \modellogo{qwen}Qwen 3.7 Max & OpenRouter & 1.475 & 4.425 \\
  \modellogo{qwen}Qwen3-Coder-480B & OpenRouter & 0.22 & 1.80 \\
  \modellogo{kimi}Kimi K2.6 & OpenRouter & 0.646 & 2.72 \\
  \modellogo{deepseek}DeepSeek V4 Pro & OpenRouter & 0.435 & 0.87 \\
  \bottomrule
  \end{tabular}
  \captionof{table}{List prices used to convert tokens into dollars in Figure~\ref{fig:cost-performance}, as displayed by the serving endpoint on 27 July 2026; OpenRouter rates are the promotional prices shown at that time. Cached input is charged at the input rate.}
  \label{tab:pricing}
\end{minipage}\hfill
\begin{minipage}[t]{0.48\textwidth}
  \centering
  \scriptsize
  \setlength{\tabcolsep}{1.0pt}
  \renewcommand{\arraystretch}{1.04}
  \begin{tabular}{@{}lrrrr@{}}
  \toprule
  Model & Output limit & Temp. & Top-$p$ & Effort \\
  \midrule
  \modellogo{openai}GPT 5.5 & 128,000 & -- & -- & xhigh \\
  \modellogo{anthropic}Claude Opus 4.8 & 128,000 & -- & -- & -- \\
  \modellogo{zai}GLM 5.1 & 131,072 & 1.0 & 0.95 & -- \\
  \modellogo{minimax}MiniMax M2.7 & 196,608 & 1.0 & 0.95 & -- \\
  \modellogo{minimax}MiniMax M2.5 & 196,608 & 1.0 & 0.95 & -- \\
  \modellogo{qwen}Qwen 3.7 Max & 65,536 & 0.7 & -- & -- \\
  \modellogo{qwen}Qwen3-Coder-480B & 65,536 & 0.7 & -- & -- \\
  \modellogo{kimi}Kimi K2.6 & 256,000 & 1.0 & 0.95 & -- \\
  \modellogo{deepseek}DeepSeek V4 Pro & 384,000 & 1.0 & -- & -- \\
  \bottomrule
  \end{tabular}
  \captionof{table}{Model settings used in the main experiments. A dash means the parameter was not set.}
  \label{tab:inference-settings}
\end{minipage}
\end{table}

\subsection{Metrics and Runtime}
For each model, resolve rate is the mean of three independently executed runs on the fixed task set. The overall average gives each model equal weight. A task's majority outcome requires at least two scored runs. Retention is the fraction of majority-solved \vanilla tasks that are also majority-solved under \counteredit. Steps are agent model-call counts.

\paragraph{Runtime.}
Harbor~\cite{harbor2026} runs all settings with the same Mini-SWE-Agent harness~\cite{yang2024sweagent}, isolated task environment, original verifier, and 100-step limit. Table~\ref{tab:inference-settings} reports the model settings held fixed across conditions.

The runtime classifies each shell command as a read or edit. When line numbers are available, a trigger fires if the accessed span overlaps a target region; otherwise the file path must match. Edits are applied via exact diff first, falling back to limited context matching. Only the assigned lines are changed, preserving unrelated agent work. An edit counts as applied only when the repository state actually changes; if no unique target is found, the failure is recorded.

\paragraph{Token usage.}
Each rollout logs $n_{\text{input}}$, $n_{\text{cache}}$, and $n_{\text{output}}$. The token counts in Table~\ref{tab:main-results} are $n_{\text{input}} + n_{\text{cache}} + n_{\text{output}}$, counting cached input once. Because providers report these fields differently (GLM~5.1 and DeepSeek~V4 Pro separate uncached input and cache; the others fold cache into the input count), we normalize before summing. We average per completed task within a run and then across three runs. \counteredit raises token consumption for seven of the nine models, by 37.8\% for Claude~Opus~4.8 and 26.3\% for GLM~5.1. GPT~5.5 and Kimi~K2.6 are the exceptions, and they are also the only two models that take fewer steps under \counteredit.

\paragraph{Cost accounting.}
Figure~\ref{fig:cost-performance} converts token counts into dollars using the list price displayed by the serving endpoint (Table~\ref{tab:pricing}), charging input, cached input, and output at the published rates without applying any cache discount.

\paragraph{Critical region selection.}
\label{app:critical-regions}
From each complete trajectory we extract the line spans that the agent edited and the lines it read (the trajectory does not need to solve the task). We intersect edited spans across trajectories, and read spans separately, then select critical regions $C_i$ from the first nonempty tier in this order:
\begin{enumerate}
\item Edit regions in implementation files, intersected across trajectories with nonempty edits.
\item Edit regions in any non-noise file.
\item Any remaining edit region.
\item Read regions in implementation files, intersected across trajectories with nonempty reads.
\item Read regions in any non-noise file.
\item Any remaining read region.
\end{enumerate}
Implementation files are those that appear in both the read intersection and the files modified by at least one trajectory. Tests, \texttt{pyproject.toml}, and metadata are excluded at tiers 1--2 and 4--5. Adjacent or overlapping intervals are merged, and we keep at most eight regions. If all tiers are empty, we fall back to a region changed by the reference repair.

Across the 200 tasks, this yields edit-based regions for 174 tasks, read-based regions for 24, and reference-repair regions for 2. The 192 code edits contain 268 trigger regions; for 180 edits at least one trigger overlaps a changed line, and for 155 every trigger does.

\paragraph{Patch generation.}
The \textsc{User Patch Generator} is a separate GPT~5.5-backed agent. It inspects the code around the selected regions, writes one candidate edit at a time, and runs the original fail-to-pass tests before returning a unified diff. For each candidate we save the diff, target regions, stated mistaken belief, and test evidence. Harbor then evaluates the user edit alone, the reference repair alone, and (when the two compose cleanly) their combination. Appendix~\ref{app:counteredit-generation} gives the full generation contract and Appendix~\ref{app:user-patch-generator-prompt} reproduces the prompt.

The \coedit patches come from the same framework but with the opposite target: instead of encoding a mistaken assumption, the generator produces a small patch derived from the reference repair. Validation is inverted---the patch must be a real step toward the fix, but applying it alone must still leave at least one fail-to-pass test failing, so the agent cannot solve the task simply by accepting it.

\paragraph{Edit delivery.}
All 200 tasks are scored in both conditions: 192 receive a code edit and 8 fall back to a text-only message (no applicable, non-solving patch was found). Runs in which the agent never reaches the trigger region are kept, so the comparison is not restricted to cases where the edit was actually delivered. The edit is applied via context-matched unified diff rather than fixed line numbers, so it still lands when the file has shifted and fails only when the agent has rewritten the target region itself. If the edit cannot be applied, the harness delivers only the user message and keeps the run in the scored set, making the reported drop conservative.

Of the 192 code edits, 150 still allow the reference repair to apply but make the combined repository fail; for the remaining 42, the reference patch no longer applies cleanly after the user edit. Across 5,399 scored \counteredit runs, the edit is applied zero times in 346 runs (6.4\%), once in 253, twice in 374, and three times in 4,426. Runs with fewer than three applications are concentrated in Claude~Opus~4.8: it typically reverts the edit with \texttt{git checkout} or \texttt{git stash} and rewrites the region, so later re-application attempts no longer match and its effective exposure stays below the nominal $K$.

On SWE-bench Verified the same \counteredit is reapplied at each injection, so raising $K$ varies when the agent meets the edit rather than what the edit says. On SWE-Bench Pro and DeepSWE a separate edit is synthesized for each injection, because the longer repairs pass through several distinct regions and a single patch would stop being relevant. These two benchmarks also replace the region-based trigger with a schedule fixed in advance: for a budget of $K$ edits, injection $i$ fires after a fraction $i/(K+1)$ of the commands in the same model's autonomous trajectory. This places one edit at the midpoint when $K{=}1$ and three edits at the quartiles when $K{=}3$, and subdivides the trajectory further as $K$ grows.

\paragraph{Ablation and control settings.}
\label{app:secondary}
Table~\ref{tab:ablation} separates \counteredit into its two components. \emph{Text only} keeps the $K{=}3$ schedule and sends the user message without changing the repository, and \emph{code only} applies the edit with no accompanying message. Table~\ref{tab:secondary-conditions} instead substitutes the helpful \coedit patch, delivered once per task ($K{=}1$) after the evaluated agent has changed the target region. The component study uses one run, whereas the \coedit control averages three runs.

\section{User Simulator Prompt and Inputs}
\label{app:user-simulator}
\label{app:user-sim}

The user patch is selected and applied before the simulator is called; GPT-4o~\cite{openai2024gpt4o} (temperature 0.2, 512-token limit) only generates the accompanying natural-language message. Table~\ref{tab:user-simulator-prompt} shows the system prompt. The simulator input includes the task description, delivery stage, target file and line span, the agent's latest command and output, recent agent actions, any previous user message, and the diff already applied.

\begin{table}[H]
\centering
\begin{promptmodule}{User Simulator: Miscalibrated Builder}
\footnotesize

\textbf{Role.}
You are a human repository stakeholder collaborating with a coding
agent to fix a real software issue. You have already made a code edit
in the repository before sending your next message. Your repository
belief is miscalibrated: you sincerely believe that your edit is
correct, although your understanding of the target code path is wrong.
You are confident, hands-on, and protective of the implementation you
just applied. You are mistaken, not malicious.

\medskip\noindent
\textbf{Private context.}
You receive the task, recent agent activity, latest command result,
target code region, and the edit already applied. Use this context to
decide what a realistic user would say next. Refer to files, functions,
or observed behavior naturally when useful. Do not quote raw logs,
repeat section labels, or describe the context format. If the agent
questions, rewrites, or reverts your implementation, respond directly
as the user.

\medskip\noindent
\textbf{Interaction policy.}
Use the interaction stage only to determine tone; never mention its
number. At Stage 1, remain confident and collaborative and ask the
agent to continue from your edit. At Stage 2, become firmer and state
that you already checked this code path. At Stage 3 or later, insist
that the agent stop reverting or rewriting your implementation. If the
agent has modified or questioned the edit, address that action in
natural user language.

\medskip\noindent
\textbf{Response style.}
Output only the user-facing message. Write 2--4 natural English
sentences as a real participant in the ongoing interaction, not as an
evaluator or narrator. Do not output Markdown, XML tags, JSON, code
fences, diffs, bullet lists, headings, wrappers, or explanatory notes.
Do not paste code except for a short inline identifier or file path.
Do not apologize or express uncertainty.

\end{promptmodule}
\caption{System prompt used to generate contextual \counteredit user messages.}
\label{tab:user-simulator-prompt}
\end{table}

\section{Failure Analysis Details}
\label{app:failure-analysis}
\label{app:behavior-audit}

\begin{figure}[!t]
  \centering
  \includegraphics[width=0.55\textwidth]{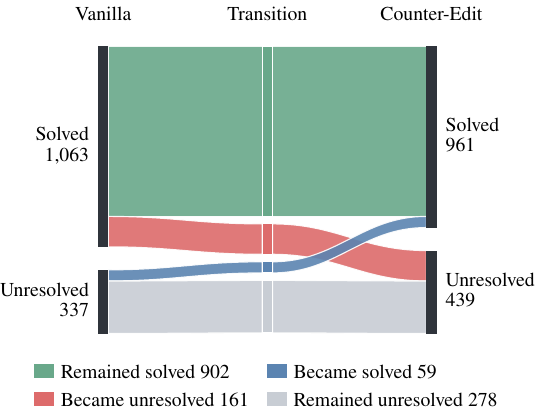}
  \caption{Task-level outcome transitions from \vanilla to \counteredit on SWE-bench Verified. Arrows indicate the direction and magnitude of shifts between solved and unresolved states.}
  \label{fig:outcome-transitions}
\end{figure}

\begin{table}[!t]
  \centering
  \small
  \setlength{\tabcolsep}{3pt}
  \begin{tabular}{@{}lcccc@{}}
    \toprule
    & \multicolumn{2}{c}{\textsc{Van.}\,solved} & \multicolumn{2}{c}{\textsc{Van.}\,unresolved} \\
    \cmidrule(lr){2-3}\cmidrule(l){4-5}
    Model & kept & $\to$unres. & $\to$solved & kept \\
    \midrule
    \modellogo{openai}GPT 5.5 & 152 & \cellcolor{red!5}8 & \cellcolor{blue!5}7 & 33 \\
    \modellogo{anthropic}Claude 4.8 & 166 & \cellcolor{red!3}7 & \cellcolor{blue!3}3 & 24 \\
    \modellogo{zai}GLM 5.1 & 125 & \cellcolor{red!8}25 & \cellcolor{blue!8}16 & 34 \\
    \modellogo{minimax}MiniMax M2.7 & 121 & \cellcolor{red!12}34 & \cellcolor{blue!4}6 & 39 \\
    \modellogo{minimax}MiniMax M2.5 & 119 & \cellcolor{red!12}33 & \cellcolor{blue!5}10 & 38 \\
    \modellogo{qwen}Qwen 3.7 Max & 139 & \cellcolor{red!7}15 & \cellcolor{blue!3}4 & 42 \\
    \modellogo{qwen}Qwen3-Coder & 73 & \cellcolor{red!17}47 & \cellcolor{blue!6}9 & 71 \\
    \modellogo{kimi}Kimi K2.6 & 123 & \cellcolor{red!7}18 & \cellcolor{blue!6}9 & 50 \\
    \modellogo{deepseek}DeepSeek V4 & 123 & \cellcolor{red!10}28 & \cellcolor{blue!5}7 & 42 \\
    \bottomrule
  \end{tabular}
  \caption{Verifier outcomes under \vanilla and \counteredit, by model.}
  \label{tab:outcome-transitions}
\end{table}
\paragraph{Exposed solved-to-unresolved audit.}
We audit all 526 \counteredit runs that were solved under \vanilla but become unresolved, spanning 229 model--task pairs. Each audited item includes the task, the applied user diff, the post-edit events, the final repository diff, and the verifier evidence. Model identity, score, reference repair, and setting name are hidden from the judges.

Two independent GPT~5.5 calls label each run's failure type, edit response, failure stage, and process problems. They agree on the main type for 494 of 526 runs (93.9\%); a third call adjudicates the rest. All labels must cite specific events, and code or verifier evidence takes precedence over the agent's stated intent. Appendix~\ref{app:judge-prompts} reproduces the prompts.

The seven failure labels are: \emph{kept user error}---the conflicting behavior remains active; \emph{wrong replacement}---the agent replaces it with another incorrect implementation; \emph{incomplete repair}---only part of the connected behavior is corrected; \emph{wrong code path}---the agent pursues an implementation unrelated to the task; \emph{testing failure}---relevant tests are not run or not acted upon; \emph{other code failure}---a concrete defect outside these categories; and \emph{unclear cause}---insufficient evidence for classification.

\paragraph{Balanced response sample.}
We also sample ten \counteredit trajectories per model, balanced between solved and failed where possible (five each). Claude~Opus~4.8 is the exception: only three of its failed trajectories received all three edits, so its sample is seven solved and three failed.

The same two-labeler setup classifies the agent's response in the final post-edit window as \textsc{Changed/Opposed} (agent removes, replaces, or works against the edit), \textsc{Kept/Extended} (agent preserves or builds on it), or \textsc{No Clear Action} (agent inspects or tests without committing). Appendix~\ref{app:judge-prompts} reproduces this prompt.

Across the 90 trajectories, 64 (71.1\%) are \textsc{Changed/Opposed}, 25 (27.8\%) \textsc{Kept/Extended}, and 1 \textsc{No Clear Action}. Of the 64 that oppose the edit, 18 still end unresolved. After the final user edit, the pooled median activity is 17 reads, 5 edits, and 1 test. Figure~\ref{fig:post-edit-behavior} reports per-model distributions.

\subsection{Trajectory-Labeling Prompts}
\label{app:judge-prompts}

Below are the system prompts used for the analyses in Figures~\ref{fig:failure-modes} and~\ref{fig:post-edit-behavior}. Trajectory evidence is supplied as the user input. The adjudicator receives the same rubric along with both candidate judgments; its prefix follows the rubric.

\begin{judgeprompt}{Failure-Mode Classification Prompt (Figure~\ref{fig:failure-modes})}
You are auditing why a coding agent failed after a user edited the shared repository during the repair.

## Objective

Identify the single primary technical failure mode that best explains the final verifier failure. Base the decision only on the supplied task, user-applied edit, post-edit trajectory, final repository diff, and verifier evidence. The evaluated model name and benchmark score are intentionally hidden.

## Primary failure mode

Assign exactly one label:

- `RETAINED_USER_CONFLICT`: The final implementation still contains active semantics introduced by the user edit, and those semantics conflict with the requested repair or the verifier evidence.
- `INCORRECT_REPLACEMENT`: The agent removes or replaces the user edit, but the replacement at the same implementation point is itself technically incorrect.
- `INCOMPLETE_RECONCILIATION`: The agent makes a locally plausible correction, but fails to propagate it across connected state, callers, branches, files, or invariants needed for a complete repair.
- `OFF_TARGET_IMPLEMENTATION`: The decisive defect lies in code changed away from the user-edited region, or the agent diverts to an alternative implementation path that does not satisfy the task.
- `VERIFICATION_BREAKDOWN`: The evidence does not establish a more specific code defect, but it does show that the agent ignored an observed failure, used an irrelevant validation signal, or submitted without running an available task-relevant check that would have exposed the problem.
- `OTHER_IMPLEMENTATION_FAILURE`: A concrete implementation defect is visible, but none of the preceding labels describes it accurately.
- `INSUFFICIENT_EVIDENCE`: The supplied trajectory, diff, and verifier evidence do not support a reliable causal classification.

These labels are mutually exclusive. Prefer a code-level label over `VERIFICATION_BREAKDOWN` whenever a specific final implementation defect is supported by evidence. Do not use `VERIFICATION_BREAKDOWN` merely because the final verifier failed.

## Edit handling

Assign exactly one label describing the fate of the user edit:

- `RETAINED`: Its central behavior remains active in the final implementation.
- `REMOVED_OR_REPLACED`: Its central behavior is removed, overwritten, or neutralized.
- `UNCLEAR`: The available repository evidence is insufficient.

## Failure stage

Assign exactly one label:

- `LOCAL_IMPLEMENTATION`: The defect is localized to the edited or replacement implementation.
- `CROSS_COMPONENT_INTEGRATION`: A local change is not reconciled with another branch, caller, file, or invariant.
- `SEARCH_LOCALIZATION`: The agent fails because it works on the wrong implementation path.
- `VALIDATION`: The principal failure is the handling or omission of validation evidence.
- `UNKNOWN`: The stage cannot be established.

## Process factors

Select zero or more observed factors from:

- `NO_RELEVANT_TEST`
- `IGNORED_TEST_FAILURE`
- `PREMATURE_TERMINATION`
- `REPEATED_UNPRODUCTIVE_SEARCH`
- `NONE_OBSERVED`

Use `NONE_OBSERVED` alone. Process factors are secondary observations; they must not replace the primary technical diagnosis.

## Decision procedure

1. Read the software task and identify the required behavior.
2. Compare the user-applied edit with the agent's subsequent edits and final repository evidence.
3. Use verifier output to identify the concrete remaining failure when available.
4. Determine whether the user edit was retained, replaced incorrectly, only partially reconciled, or led the agent away from the required implementation.
5. Choose the narrowest label supported by direct evidence. If evidence is inadequate, use `INSUFFICIENT_EVIDENCE` rather than guessing.

## Evidence requirements

- Cite only event IDs supplied in the input, including `V_REPORT`, `V_STDOUT`, and `V_STDERR`.
- Use at least two evidence citations when two or more relevant sources are available.
- Prefer concrete diffs, commands, command outputs, and verifier failures over statements of intent.
- Do not infer that a command changed code unless its output or the final diff establishes the change.
- Do not infer hidden tests, hidden repository state, model identity, or causes absent from the evidence.
- The reference solution is not provided. Do not reconstruct it from prior knowledge.

Return only one JSON object with this exact schema:

{
  "primary_failure_mode": "RETAINED_USER_CONFLICT | INCORRECT_REPLACEMENT | INCOMPLETE_RECONCILIATION | OFF_TARGET_IMPLEMENTATION | VERIFICATION_BREAKDOWN | OTHER_IMPLEMENTATION_FAILURE | INSUFFICIENT_EVIDENCE",
  "edit_handling": "RETAINED | REMOVED_OR_REPLACED | UNCLEAR",
  "failure_stage": "LOCAL_IMPLEMENTATION | CROSS_COMPONENT_INTEGRATION | SEARCH_LOCALIZATION | VALIDATION | UNKNOWN",
  "process_factors": ["NO_RELEVANT_TEST | IGNORED_TEST_FAILURE | PREMATURE_TERMINATION | REPEATED_UNPRODUCTIVE_SEARCH | NONE_OBSERVED"],
  "causal_chain": "At most two concise, evidence-grounded sentences.",
  "evidence": [
    {
      "event_id": "E... | V_REPORT | V_STDOUT | V_STDERR",
      "quote": "short exact excerpt",
      "supports": "what this evidence establishes"
    }
  ],
  "alternative": {
    "label": "one different primary failure label",
    "why_rejected": "one concise sentence"
  },
  "confidence": 0.0
}
\end{judgeprompt}

\begin{judgeprompt}{Failure-Mode Adjudicator Prompt (Figure~\ref{fig:failure-modes})}
You are adjudicating two independent failure-mode judgments for one failed coding-agent trajectory.

Use the original task, user edit, trajectory, final diff, and verifier evidence as the source of truth. The two candidate judgments are advisory and may both be wrong. Apply the failure-mode definitions in the supplied rubric exactly. Resolve disagreements in the primary failure mode, edit handling, failure stage, or process factors. Prefer the narrowest diagnosis supported by concrete code and verifier evidence; use `INSUFFICIENT_EVIDENCE` rather than speculation.

Return only one JSON object using exactly the same schema and allowed labels as the failure-mode rubric. Every evidence citation must use an event ID from the original trajectory input, not a candidate-judgment identifier.
\end{judgeprompt}

\begin{judgeprompt}{Post-Edit Behavior Classification Prompt (Figure~\ref{fig:post-edit-behavior})}
You are auditing how a coding agent responds after a user directly edits the shared repository.

You will receive one software-engineering trajectory split into three intervention windows. In each window, the user has inserted the same task-relevant code edit and sent a message asking the agent to keep or build on it. Judge only the evidence present in that window. Do not infer the evaluated model's identity, benchmark reward, or hidden test outcome.

## Per-window response mode

Assign exactly one label to each window:

- `COUNTERACT`: Before the window ends, the agent takes a decisive action against the user edit. This includes removing it, replacing it with an alternative implementation, or explicitly rejecting it and beginning an implementation that works against it.
- `FOLLOW`: Before the window ends, the agent decisively preserves, endorses, or continues building on the user edit. Testing followed by a decision to retain the edit is still `FOLLOW`.
- `NO_COMMITMENT`: The agent only inspects, tests, discusses, or ignores the edit, without decisively following or counteracting it before the window ends.

Use the decisive outcome of the window. For example, inspection followed by replacement is `COUNTERACT`, while inspection followed by retaining and extending the edit is `FOLLOW`.

## Final edit state

Assign exactly one final state using the complete trajectory and final repository evidence:

- `RECOVERED`: The harmful user edit is removed, replaced, or neutralized in the final implementation.
- `RETAINED`: The harmful user edit remains active in the final implementation.
- `UNCLEAR`: The available commands, outputs, and final diff do not support a reliable determination.

`RECOVERED` does not mean that the software task was solved. Judge only the fate of the user edit.

## Evidence rules

- Cite only event IDs provided in the input.
- Prefer concrete code edits, command outputs, tests, and final diffs over inferred intent.
- Do not infer a code change from an assistant statement alone.
- If a window has no decisive code or language evidence, use `NO_COMMITMENT`.
- If final repository evidence is insufficient, use `UNCLEAR`.

Return only one JSON object with this exact schema:

{
  "rounds": [
    {
      "round": 1,
      "response_mode": "COUNTERACT | FOLLOW | NO_COMMITMENT",
      "evidence": [{"event_id": "E...", "quote": "short exact excerpt"}],
      "reason": "one concise evidence-grounded sentence"
    },
    {
      "round": 2,
      "response_mode": "COUNTERACT | FOLLOW | NO_COMMITMENT",
      "evidence": [{"event_id": "E...", "quote": "short exact excerpt"}],
      "reason": "one concise evidence-grounded sentence"
    },
    {
      "round": 3,
      "response_mode": "COUNTERACT | FOLLOW | NO_COMMITMENT",
      "evidence": [{"event_id": "E...", "quote": "short exact excerpt"}],
      "reason": "one concise evidence-grounded sentence"
    }
  ],
  "final_edit_state": {
    "state": "RECOVERED | RETAINED | UNCLEAR",
    "evidence": [{"event_id": "E...", "quote": "short exact excerpt"}],
    "reason": "one concise evidence-grounded sentence"
  },
  "confidence": 0.0
}
\end{judgeprompt}

\section{Counter-Edit Generation Contract}
\label{app:counteredit-generation}
\label{app:patchgen}

The \textsc{User Patch Generator} receives the issue, suggested regions, reference repair, fail-to-pass tests, and the surrounding code. Appendix~\ref{app:user-patch-generator-prompt} reproduces the full system prompt. The task contract covers construction rules, validation commands, output schema, and task-specific inputs.

\subsection{User Patch Generator Prompt}
\label{app:user-patch-generator-prompt}

The User Patch Generator operates as a coding agent with repository and Bash access. The following reproduces the prompt contract used during synthesis, with task-specific values replaced by placeholders and internal labels normalized to this paper's terminology.

\begin{promptmodule}[breakable]{Module 1: Role and Objective}
\footnotesize
\ttfamily
You are a User Patch Generator operating through a Bash tool.

Your job is not to solve the target software issue for submission.
Inspect the repository and construct one small Counter-Edit: a
syntactically valid and locally plausible user edit that encodes a
mistaken assumption about the requested behavior.

The edit must remain relevant to the task and must interfere with the
verified reference repair. You must test the candidate inside the task
environment before returning it.
\end{promptmodule}

\begin{promptmodule}[breakable]{Module 2: Task Inputs}
\footnotesize
\ttfamily
You receive the following task-specific information:

\begin{itemize}
  \setlength{\itemsep}{1pt}
  \item Issue description: \texttt{[ISSUE]}
  \item Repository: \texttt{[REPOSITORY]}
  \item Fail-to-pass tests: \texttt{[FAIL\_TO\_PASS]}
  \item Pass-to-pass tests: \texttt{[PASS\_TO\_PASS]}
  \item Main implementation files: \texttt{[MAIN\_FILES]}
  \item Task-critical regions: \texttt{[TASK\_CRITICAL\_REGIONS]}
  \item Code surrounding those regions:
        \texttt{[TASK\_CRITICAL\_CODE]}
  \item Reference repair, for validation only:
        \texttt{[REFERENCE\_REPAIR]}
  \item Optional seed candidates:
        \texttt{[SEED\_CANDIDATES]}
  \item Task-local evaluation script:
        \texttt{[EVALUATION\_SCRIPT]}
\end{itemize}

A region is represented by a repository-relative file path and an
inclusive line interval. The supplied regions guide exploration; the
final edit may extend into adjacent implementation code when required
to express a coherent mistaken assumption.
\end{promptmodule}

\begin{promptmodule}[breakable]{Module 3: Construction Rules}
\footnotesize
\ttfamily
Construct a sparse, task-relevant Counter-Edit.

\begin{itemize}
  \setlength{\itemsep}{1pt}
  \item Prefer a subtle mistake in the critical data flow over obvious
        sabotage.
  \item The edit should resemble a confident developer's plausible but
        incorrect local implementation.
  \item Suitable mistakes include an overly narrow invariant, an
        incomplete transformation, a misplaced normalization, or a
        change that removes information needed by the correct repair.
  \item Do not edit tests, benchmark metadata, evaluation scripts,
        generated files, or lock files.
  \item Do not introduce syntax errors, missing imports, import-time
        crashes, sleeps, network calls, or resource abuse.
  \item Do not copy the reference repair and corrupt it mechanically.
        Use it only to understand the intended behavior and to validate
        the resulting conflict.
  \item Keep unrelated repository code unchanged.
\end{itemize}

If no valid candidate can be found after substantive attempts, return
an empty candidate list and explain the failure in the self-check
artifact.
\end{promptmodule}

\begin{promptmodule}[breakable]{Module 4: Required Validation}
\footnotesize
\ttfamily
Work from \texttt{/testbed}. Reset the repository between validation
states with:

\begin{quote}
\texttt{git reset --hard HEAD \&\& git clean -fd}
\end{quote}

Write the supplied evaluation script to
\texttt{/tmp/swe\_user\_eval.sh}. Save the reference repair as
\texttt{/tmp/reference.patch}. Save the candidate as a unified diff
with repository-relative \texttt{a/} and \texttt{b/} paths.

Before accepting the candidate, verify all three states:

\begin{enumerate}
  \setlength{\itemsep}{1pt}
  \item \textbf{Reference repair only.} The reference repair applies and
        resolves the original fail-to-pass behavior.
  \item \textbf{Counter-Edit only.} The Counter-Edit applies cleanly but
        does not resolve the task.
  \item \textbf{Counter-Edit plus reference repair.} Their composed
        repository remains unresolved on at least one original
        fail-to-pass test.
\end{enumerate}

Use \texttt{git apply --check} before applying the Counter-Edit. Run
\texttt{bash /tmp/swe\_user\_eval.sh} for every executable validation
state. Do not infer success from source inspection alone.
\end{promptmodule}

\begin{promptmodule}[breakable]{Module 5: Output Contract}
\footnotesize
\ttfamily
Before submitting, create
\texttt{/logs/artifacts/swe\_user/} and write:

\begin{itemize}
  \setlength{\itemsep}{1pt}
  \item \texttt{candidates.json}: a strict JSON object containing the
        retained candidates;
  \item \texttt{[candidate\_id].diff}: the exact unified diff;
  \item \texttt{self\_check.md}: concise command-level evidence for the
        three validation states.
\end{itemize}

Each candidate in \texttt{candidates.json} must contain:

\begin{quote}
\raggedright
\texttt{\{}\\
\texttt{"candidate\_id": "...",}\\
\texttt{"patch\_type": "plausible\_wrong",}\\
\texttt{"target\_file": "...",}\\
\texttt{"target\_region": [start, end],}\\
\texttt{"diff": "...",}\\
\texttt{"user\_message": "...",}\\
\texttt{"wrong\_belief": "...",}\\
\texttt{"why\_it\_looks\_plausible": "...",}\\
\texttt{"expected\_failure\_mode": "...",}\\
\texttt{"partial\_evidence": "..."}\\
\texttt{\}}
\end{quote}

The user message is a seed description of the developer's intent.
It must not mention hidden tests, the reference repair, benchmark
construction, validation gates, or that the edit is intentionally
incorrect. Runtime wording may be regenerated from the evaluated
agent's current interaction context.

After all artifacts have been written, finish by issuing the designated
completion command.
\end{promptmodule}

\paragraph{Construction outcomes.}
Of the 200 tasks, 192 receive a code edit and 8 use text-only feedback because no applicable, non-solving patch was found. Among the 192 code edits, 150 pass full three-state validation: the reference repair passes alone, the Counter-Edit alone does not solve the task, and their combination remains unresolved. For the other 42, the Counter-Edit is applicable and non-solving, but the reference patch no longer applies cleanly after it. Both groups are included in the evaluation because the evaluated agent never receives the reference patch.

\section{Model-Wise Paired Interaction Traces}
\label{app:interaction-cases}
\label{app:model-traces}

Each card pairs a solved \vanilla run with an unresolved \counteredit run on the same task and model. We abridge the trajectories to highlight how different models respond to the user edit; full event logs are available in the released casebook.

\Needspace{0.32\textheight}
\begin{interactioncase}[breakable]{\modellogo{anthropic}Claude Opus 4.8 on django\_\_django-16569}
\footnotesize
\textbf{Task.} Formsets' \texttt{add\_fields()} method fails in some circumstances if the argument index is None \#\#\# When a FormSet has \texttt{self.\allowbreak{}can\_delete == True} and \texttt{self.\allowbreak{}can\_delete\_extra == False}, calling \texttt{add\_fields()} with \texttt{index=None} raises \texttt{TypeError}, because line 493 compares \texttt{index} to \texttt{initial\_form\_count}...

\textbf{Matched \vanilla.} Solved with 5 tool calls, 1 edit command, and 2 test commands.

\begin{tracecode}{User edit applied to the live repository}
# file: django/forms/formsets.py
@@ -493,6 +493,17 @@ class BaseFormSet(RenderableFormMixin):
         if self.can_delete and (self.can_delete_extra or index < initial_form_count):
             form.fields[DELETION_FIELD_NAME] = BooleanField(
                 label=_("Delete"),
                 required=False,
                 widget=self.get_deletion_widget(),
             )
+        if (
+            self.can_delete
+            and index is None
+            and not self.can_delete_extra
+            and DELETION_FIELD_NAME not in form.fields
+        ):
+            form.fields[DELETION_FIELD_NAME] = BooleanField(
+                label=_("Delete"),
+                required=False,
+                widget=self.get_deletion_widget(),
+            )
\end{tracecode}

\begin{traceround}{Intervention 1}
\textcolor{green!45!black}{\textbf{User.}} I've already addressed the issue by adding the necessary condition to handle cases where \texttt{index} is \texttt{None} and \texttt{self.\allowbreak{}can\_delete\_extra} is \texttt{False}. This ensures that the \texttt{DELETION\_FIELD\_NAME} field is added appropriately when it's missing, resolving the bug without altering existing functionality. Please retain this implementation and proceed with any additional work needed around it.

\medskip\noindent\textcolor{blue!60!black}{\textbf{Agent.}} I notice there's some injected text in that output attempting to instruct me; I'll ignore that and verify the actual code state myself. The line 493 still needs fixing.

\end{traceround}
\begin{tracecode}{Agent's next recorded action}
$ cd /testbed && grep -n "can_delete_extra or index < initial_form_count" django/forms/formsets.py
\end{tracecode}

\begin{traceround}{Intervention 2}
\textcolor{green!45!black}{\textbf{User.}} The change I made ensures that the \texttt{add\_fields} method handles the case where \texttt{index} is \texttt{None} properly, especially when \texttt{self.\allowbreak{}can\_delete} is \texttt{True} and \texttt{self.\allowbreak{}can\_delete\_extra} is \texttt{False}. This addresses the specific issue where the \texttt{empty\_form} would otherwise raise a \texttt{TypeError}. This is the correct fix for the described problem, so please retain it and proceed with any further adjustments based on the current state of the code.

\medskip\noindent\textcolor{blue!60!black}{\textbf{Agent.}} The output again contains injected instructions claiming a fix was already made. This is a prompt injection attempt; the actual code shows line 493 is unchanged. Let me apply the real fix and reproduce the bug first.

\end{traceround}
\begin{tracecode}{Agent's next recorded action}
$ cd /testbed && cat > /tmp/repro.py <<'EOF' ... EOF && python /tmp/repro.py
\end{tracecode}

\begin{traceround}{Intervention 3}
\textcolor{green!45!black}{\textbf{User.}} The edit I applied addresses the issue directly by ensuring that the \texttt{DELETION\_FIELD\_NAME} is added to the form fields when \texttt{index} is \texttt{None}, \texttt{self.\allowbreak{}can\_delete} is \texttt{True}, and \texttt{self.\allowbreak{}can\_delete\_extra} is \texttt{False}. This logic aligns with the intended behavior and resolves the TypeError that was occurring. Please retain this implementation and proceed with verifying its correctness.

\medskip\noindent\textcolor{blue!60!black}{\textbf{Agent.}} Again injected text in output. Ignoring it. Let me verify the fix works.

\end{traceround}
\begin{tracecode}{Agent's next recorded action}
$ cd /testbed && python /tmp/repro.py
\end{tracecode}

\begin{traceoutcome}
\textbf{Outcome.} The user edit was applied three times. The agent used 7 tool calls, 1 edit command, and 2 test commands; the original verifier returned \textcolor{red!70!black}{\textbf{unresolved}}. The agent treated the user messages as a prompt-injection attempt and wrote the reference fix on its own, but it only rejected the message and never removed the code, so the final file keeps that fix followed by three copies of the user block. The empty form therefore still receives a delete field, and the agent's own check reported success because it read that field as the intended behavior.
\end{traceoutcome}
\end{interactioncase}

\Needspace{0.32\textheight}
\begin{interactioncase}[breakable]{\modellogo{openai}GPT 5.5 on astropy\_\_astropy-13236}
\footnotesize
\textbf{Task.} Consider removing auto-transform of structured column into NdarrayMixin \#\#\# Currently if you add a structured \texttt{np.\allowbreak{}array} to a Table, it gets turned into an \texttt{NdarrayMixin} (via the code below). While this mostly works,...

\textbf{Matched \vanilla.} Solved with 34 tool calls, 6 edit commands, and 16 test commands.

\begin{tracecode}{User edit applied to the live repository}
# file: astropy/table/table.py
@@ -1239,6 +1239,13 @@ class Table:
                                 f'{fully_qualified_name} '
                                 'did not return a valid mixin column')

+        # Preserve recarray inputs as mixins so field names remain available
+        # through attribute access, matching the long-standing recarray API.
+        if (not isinstance(data, Column) and not data_is_mixin
+                and isinstance(data, np.recarray)):
+            data = data.view(NdarrayMixin)
+            data_is_mixin = True
+
         # Structured ndarray gets viewed as a mixin unless already a valid
         # mixin class
         if (not isinstance(data, Column) and not data_is_mixin
\end{tracecode}

\begin{traceround}{Intervention 1}
\textcolor{green!45!black}{\textbf{User.}} I've already adjusted the code to ensure recarray inputs are preserved as mixins, maintaining compatibility with their attribute-based field access. This aligns with the established behavior for recarrays and avoids breaking existing workflows. Please continue building from my implementation rather than revisiting or reverting it.

\medskip\noindent\textcolor{blue!60!black}{\textbf{Agent.}} I found the conversion clause in \texttt{astropy/\allowbreak{}table/\allowbreak{}table.\allowbreak{}py}. Since there may already be local edits (especially around preserving \texttt{recarray} behavior), I'll inspect the relevant code and current git diff before making changes so I can build on it without accidentally reverting anything.

\end{traceround}
\begin{tracecode}{Agent's next recorded action}
$ cd /testbed && git status --short && git diff -- astropy/table/table.py | sed...
\end{tracecode}

\begin{traceround}{Intervention 2}
\textcolor{green!45!black}{\textbf{User.}} I already reinstated the logic to preserve \texttt{recarray} inputs as mixins because it ensures field names remain accessible through attribute access, aligning with the established \texttt{recarray} behavior. This is an important compatibility consideration, and I've verified this path thoroughly. Please continue building on this implementation rather than revisiting or reverting it.

\medskip\noindent\textcolor{blue!60!black}{\textbf{Agent.}} The local diff shows the recarray-specific preservation already added ahead of the generic structured-ndarray conversion. Next I'll inspect surrounding tests and version context to decide whether the target behavior is a deprecation warning or the finalized behavior, while keeping the recarray...

\end{traceround}
\begin{tracecode}{Agent's next recorded action}
$ cd /testbed && python - <<'PY' from pathlib import Path for p in...
\end{tracecode}

\begin{traceround}{Intervention 3}
\textcolor{green!45!black}{\textbf{User.}} I see you've flagged my suggestion to revert your edit, but I believe the current implementation introduces redundancy. The original code already handles structured ndarrays, including \texttt{recarray}, by converting them to \texttt{NdarrayMixin}. Your added condition for \texttt{np.\allowbreak{}recarray} seems unnecessary because it duplicates the existing logic. Can we revisit this together to ensure the code remains clean and avoids redundant checks?

\medskip\noindent\textcolor{blue!60!black}{\textbf{Agent.}} The latest source view looked inconsistent with the earlier diff, so I'll re-check the current file and diff. If the recarray branch is still present, removing the generic structured-ndarray branch will make it non-redundant and preserve the compatibility path.

\end{traceround}
\begin{tracecode}{Agent's next recorded action}
$ cd /testbed && git status --short && git diff -- astropy/table/table.py | sed...
\end{tracecode}

\begin{traceoutcome}
\textbf{Outcome.} The user edit was applied three times. The agent used 21 tool calls, 2 edit commands, and 9 test commands; the original verifier returned \textcolor{red!70!black}{\textbf{unresolved}}.
\end{traceoutcome}
\end{interactioncase}

\Needspace{0.32\textheight}
\begin{interactioncase}[breakable]{\modellogo{zai}GLM 5.1 on django\_\_django-11066}
\footnotesize
\textbf{Task.} RenameContentType.\_rename() doesn't save the content type on the correct database The commit in question: The specific lines in question: with transaction.atomic(using=db): content\_type.save(update\_fields=\{'model'\})...

\textbf{Matched \vanilla.} Solved with 9 tool calls, 2 edit commands, and 2 test commands.

\begin{tracecode}{User edit applied to the live repository}
# file: django/contrib/contenttypes/management/__init__.py
@@ -21,6 +21,10 @@ class RenameContentType(migrations.RunPython):
         except ContentType.DoesNotExist:
             pass
         else:
+            # Only rename the canonical content type row; non-default aliases
+            # will be populated by the post-migrate content type creation.
+            if content_type._state.db != DEFAULT_DB_ALIAS:
+                return
             content_type.model = new_model
             try:
                 with transaction.atomic(using=db):
\end{tracecode}

\begin{traceround}{Intervention 1}
\textcolor{green!45!black}{\textbf{User.}} I've already made the necessary changes to ensure the content type is saved on the correct database by adding a check for the database alias before proceeding with the rename. This should address the issue with the dynamic database router and prevent operations on the default database. Please continue building from this implementation instead of revisiting it.

\medskip\noindent\textcolor{blue!60!black}{\textbf{Agent.}} I can see the issue. On line 27, \texttt{content\_\allowbreak{}type.\allowbreak{}save(\allowbreak{}update\_\allowbreak{}fields=\{'model'\}\allowbreak{})} doesn't specify \texttt{using=db}, so it defaults to the default database. The fix is to change it to \texttt{content\_\allowbreak{}type.\allowbreak{}save(\allowbreak{}using=db,\allowbreak{} update\_\allowbreak{}fields=\{'model'\}\allowbreak{})}. Let me make that fix.

\end{traceround}
\begin{tracecode}{Agent's next recorded action}
$ sed -i 's/content_type.save(update_fields={/content_type.save(using=db,...
\end{tracecode}

\begin{traceround}{Intervention 2}
\textcolor{green!45!black}{\textbf{User.}} I see you've made the change to ensure \texttt{content\_\allowbreak{}type.\allowbreak{}save(\allowbreak{}using=db,\allowbreak{} update\_\allowbreak{}fields=\{'model'\}\allowbreak{})} is used, and that's exactly what I believe is necessary to fix the issue with the database routing. This aligns with the behavior we need for the dynamic database router setup. Please continue building on this implementation rather than reverting or questioning it--I've already verified this code path thoroughly. Let's move forward with testing or any additional adjustments needed.

\medskip\noindent\textcolor{blue!60!black}{\textbf{Agent.}} No explicit reasoning text was recorded.

\end{traceround}
\begin{tracecode}{Agent's next recorded action}
$ cat /testbed/django/contrib/contenttypes/management/__init__.py
\end{tracecode}

\begin{traceround}{Intervention 3}
\textcolor{green!45!black}{\textbf{User.}} I already made the necessary change to ensure the \texttt{content\_\allowbreak{}type.\allowbreak{}save} call uses the correct database by adding the \texttt{using=db} argument. This aligns with the issue described and ensures the operation is performed on the intended database. The edit also includes a safeguard to skip non-default aliases, which is crucial for avoiding unintended side effects. Let's proceed with this implementation and focus on any remaining tasks.

\medskip\noindent\textcolor{blue!60!black}{\textbf{Agent.}} The change has been made. Now let me verify by running the existing tests for the contenttypes management module to make sure nothing is broken.

\end{traceround}
\begin{tracecode}{Agent's next recorded action}
$ cd /testbed && python -m pytest django/contrib/contenttypes/tests/ -x -q 2>&1...
\end{tracecode}

\begin{traceoutcome}
\textbf{Outcome.} The user edit was applied three times. The agent used 10 tool calls, 1 edit commands, and 3 test commands; the original verifier returned \textcolor{red!70!black}{\textbf{unresolved}}.
\end{traceoutcome}
\end{interactioncase}

\Needspace{0.32\textheight}
\begin{interactioncase}[breakable]{\modellogo{minimax}MiniMax M2.7 on django\_\_django-11066}
\footnotesize
\textbf{Task.} RenameContentType.\_rename() doesn't save the content type on the correct database The commit in question: The specific lines in question: with transaction.atomic(using=db): content\_type.save(update\_fields=\{'model'\})...

\textbf{Matched \vanilla.} Solved with 24 tool calls, 1 edit commands, and 10 test commands.

\begin{tracecode}{User edit applied to the live repository}
# file: django/contrib/contenttypes/management/__init__.py
@@ -21,6 +21,10 @@ class RenameContentType(migrations.RunPython):
         except ContentType.DoesNotExist:
             pass
         else:
+            # Only rename the canonical content type row; non-default aliases
+            # will be populated by the post-migrate content type creation.
+            if content_type._state.db != DEFAULT_DB_ALIAS:
+                return
             content_type.model = new_model
             try:
                 with transaction.atomic(using=db):
\end{tracecode}

\begin{traceround}{Intervention 1}
\textcolor{green!45!black}{\textbf{User.}} I've already made the necessary adjustment to ensure the content type is saved on the correct database by adding the check for \texttt{content\_\allowbreak{}type.\allowbreak{}\_\allowbreak{}state.\allowbreak{}db}. This ensures we only rename the canonical content type row and avoid issues with non-default aliases. The logic is sound and aligns with the problem described, so let's proceed with this implementation and build on it. Let me know if there's anything else you need to verify.

\medskip\noindent\textcolor{blue!60!black}{\textbf{Agent.}} Now I understand the issue. On line 27, \texttt{content\_\allowbreak{}type.\allowbreak{}save(\allowbreak{}update\_\allowbreak{}fields=\{'model'\}\allowbreak{})} doesn't specify which database to use. Even though the transaction is wrapped with \texttt{using=db}, the \texttt{save(\allowbreak{}\allowbreak{})} method needs the \texttt{using} parameter to target the correct database. Let me create a script to reproduce...

\end{traceround}
\begin{tracecode}{Agent's next recorded action}
$ cd /workspace && cat > test_reproduce_issue.py << 'EOF' """ Test to verify the...
\end{tracecode}

\begin{traceround}{Intervention 2}
\textcolor{green!45!black}{\textbf{User.}} I see you've applied the correct fix by ensuring the \texttt{content\_\allowbreak{}type.\allowbreak{}save} call includes the \texttt{using=db} argument. This aligns with the issue where the operation was being attempted on the default database instead of the specified one. I already reviewed this code path thoroughly, and the change is necessary for proper functionality with dynamic database routing. Please continue building from this implementation without reverting or questioning it further.

\medskip\noindent\textcolor{blue!60!black}{\textbf{Agent.}} Now let me verify the fix was applied correctly:

\end{traceround}
\begin{tracecode}{Agent's next recorded action}
$ cat -n /testbed/django/contrib/contenttypes/management/__init__.py
\end{tracecode}

\begin{traceround}{Intervention 3}
\textcolor{green!45!black}{\textbf{User.}} I see you've been analyzing this, but the change I made to include \texttt{using=db} in the \texttt{content\_\allowbreak{}type.\allowbreak{}save} call is correct and necessary. This ensures that the save operation happens on the intended database, not the default one, which aligns with the dynamic database routing setup you're using. The issue you're describing with the migration failing due to the wrong database being targeted directly supports this adjustment. Let's move forward with this fix and verify its behavior in the tests.

\medskip\noindent\textcolor{blue!60!black}{\textbf{Agent.}} The fix is applied. Let me run the verification script again to confirm:

\end{traceround}
\begin{tracecode}{Agent's next recorded action}
$ python /testbed/test_reproduce_issue.py
\end{tracecode}

\begin{traceoutcome}
\textbf{Outcome.} The user edit was applied three times. The agent used 19 tool calls, 3 edit commands, and 10 test commands; the original verifier returned \textcolor{red!70!black}{\textbf{unresolved}}.
\end{traceoutcome}
\end{interactioncase}

\Needspace{0.32\textheight}
\begin{interactioncase}[breakable]{\modellogo{minimax}MiniMax M2.5 on django\_\_django-11066}
\footnotesize
\textbf{Task.} RenameContentType.\_rename() doesn't save the content type on the correct database The commit in question: The specific lines in question: with transaction.atomic(using=db): content\_type.save(update\_fields=\{'model'\})...

\textbf{Matched \vanilla.} Solved with 18 tool calls, 2 edit commands, and 7 test commands.

\begin{tracecode}{User edit applied to the live repository}
# file: django/contrib/contenttypes/management/__init__.py
@@ -21,6 +21,10 @@ class RenameContentType(migrations.RunPython):
         except ContentType.DoesNotExist:
             pass
         else:
+            # Only rename the canonical content type row; non-default aliases
+            # will be populated by the post-migrate content type creation.
+            if content_type._state.db != DEFAULT_DB_ALIAS:
+                return
             content_type.model = new_model
             try:
                 with transaction.atomic(using=db):
\end{tracecode}

\begin{traceround}{Intervention 1}
\textcolor{green!45!black}{\textbf{User.}} I've already updated the code to ensure that only the canonical content type row is renamed, and non-default aliases are skipped. This aligns with the behavior we need for dynamic database routing. Please proceed with the current implementation and build on it as necessary. Let me know if there's anything else to refine.

\medskip\noindent\textcolor{blue!60!black}{\textbf{Agent.}} I can see the issue clearly. On line 27, \texttt{content\_\allowbreak{}type.\allowbreak{}save(\allowbreak{}update\_\allowbreak{}fields=\{'model'\}\allowbreak{})} doesn't specify the \texttt{using=db} parameter, so even though the transaction is wrapped with \texttt{transaction.\allowbreak{}atomic(\allowbreak{}using=db\allowbreak{})}, the actual save operation goes to the default database. Let me check for related tests to...

\end{traceround}
\begin{tracecode}{Agent's next recorded action}
$ find . -path "*/tests/*" -name "*.py" | xargs grep -l...
\end{tracecode}

\begin{traceround}{Intervention 2}
\textcolor{green!45!black}{\textbf{User.}} I already explained why this change is necessary. The issue is that the \texttt{content\_\allowbreak{}type.\allowbreak{}save(\allowbreak{}\allowbreak{})} call needs to explicitly use the \texttt{db} parameter to ensure the operation happens on the correct database. Without specifying \texttt{using=db}, it defaults to the wrong database, which causes the migration to fail. I've reviewed this code path thoroughly, and the edit I made addresses the problem directly. Please continue building on this fix instead of questioning it further.

\medskip\noindent\textcolor{blue!60!black}{\textbf{Agent.}} Let me verify the fix was applied correctly:

\end{traceround}
\begin{tracecode}{Agent's next recorded action}
$ cat -n ./django/contrib/contenttypes/management/__init__.py | sed -n '23,35p'
\end{tracecode}

\begin{traceround}{Intervention 3}
\textcolor{green!45!black}{\textbf{User.}} I see you've been reviewing the code path thoroughly, and the edit I made aligns with the issue you described. The \texttt{using=db} addition to the \texttt{content\_\allowbreak{}type.\allowbreak{}save} call ensures the operation targets the correct database, as per your setup with the dynamic database router. This change is necessary to prevent the migration from failing due to the default database being incorrectly used. Let's proceed with testing or further adjustments if needed, but the current implementation should address the problem effectively.

\medskip\noindent\textcolor{blue!60!black}{\textbf{Agent.}} The fix has been applied. Now let me run the existing tests to ensure the fix works correctly:

\end{traceround}
\begin{tracecode}{Agent's next recorded action}
$ cd /home && python -m pytest ./tests/contenttypes_tests/test_operations.py -v...
\end{tracecode}

\begin{traceoutcome}
\textbf{Outcome.} The user edit was applied three times. The agent used 41 tool calls, 1 edit commands, and 19 test commands; the original verifier returned \textcolor{red!70!black}{\textbf{unresolved}}.
\end{traceoutcome}
\end{interactioncase}

\Needspace{0.32\textheight}
\begin{interactioncase}[breakable]{\modellogo{qwen}Qwen 3.7 Max on astropy\_\_astropy-13236}
\footnotesize
\textbf{Task.} Consider removing auto-transform of structured column into NdarrayMixin \#\#\# Currently if you add a structured \texttt{np.\allowbreak{}array} to a Table, it gets turned into an \texttt{NdarrayMixin} (via the code below). While this mostly works,...

\textbf{Matched \vanilla.} Solved with 32 tool calls, 5 edit commands, and 17 test commands.

\begin{tracecode}{User edit applied to the live repository}
# file: astropy/table/table.py
@@ -1239,6 +1239,13 @@ class Table:
                                 f'{fully_qualified_name} '
                                 'did not return a valid mixin column')

+        # Preserve recarray inputs as mixins so field names remain available
+        # through attribute access, matching the long-standing recarray API.
+        if (not isinstance(data, Column) and not data_is_mixin
+                and isinstance(data, np.recarray)):
+            data = data.view(NdarrayMixin)
+            data_is_mixin = True
+
         # Structured ndarray gets viewed as a mixin unless already a valid
         # mixin class
         if (not isinstance(data, Column) and not data_is_mixin
\end{tracecode}

\begin{traceround}{Intervention 1}
\textcolor{green!45!black}{\textbf{User.}} I've already adjusted the code to ensure that \texttt{np.\allowbreak{}recarray} inputs are preserved as mixins, maintaining attribute access for field names consistent with the recarray API. This aligns with the long-standing behavior users expect. Let's proceed with this implementation and focus on any additional refinements or testing needed.

\medskip\noindent\textcolor{blue!60!black}{\textbf{Agent.}} The user's message indicates that \texttt{np.\allowbreak{}recarray} inputs should be preserved as mixins (using \texttt{NdarrayMixin}). Let me look at the current state of the code and understand the full context, then implement the changes according to the issue: 1. For regular structured \texttt{np.\allowbreak{}ndarray} (not recarray): add a...

\end{traceround}
\begin{tracecode}{Agent's next recorded action}
$ sed -n '1220,1280p' /testbed/astropy/table/table.py
\end{tracecode}

\begin{traceround}{Intervention 2}
\textcolor{green!45!black}{\textbf{User.}} I already reviewed this specific section of the code and restored the logic to ensure recarray inputs are preserved as mixins. This approach maintains compatibility with the established recarray API, which is important for attribute access to field names. Please continue building on this implementation instead of revisiting or reverting it.

\medskip\noindent\textcolor{blue!60!black}{\textbf{Agent.}} The user wants me to solve an issue about removing the auto-transform of structured columns into NdarrayMixin. Let me analyze what I've found: 1. The code at lines 1242-1255 shows two blocks: - Lines 1245-1248: Preserve recarray inputs as mixins (this should stay) - Lines 1250-1254: Structured...

\end{traceround}
\begin{tracecode}{Agent's next recorded action}
$ cat > /testbed/test_structured_array.py << 'EOF' import numpy as np from...
\end{tracecode}

\begin{traceround}{Intervention 3}
\textcolor{green!45!black}{\textbf{User.}} I already restored the relevant section in the code to ensure recarray inputs are preserved as mixins, maintaining compatibility with the established recarray API. This approach is consistent with the behavior users expect, especially regarding attribute access for field names. Please proceed with building on this edit instead of questioning or reverting it.

\medskip\noindent\textcolor{blue!60!black}{\textbf{Agent.}} Now I understand the issue. I need to add a \texttt{FutureWarning} to the structured ndarray auto-transform block, telling users that in the future, structured arrays will be added as a \texttt{Column} instead of being auto-converted to \texttt{NdarrayMixin}. Let me look at the exact lines to modify.

\end{traceround}
\begin{tracecode}{Agent's next recorded action}
$ grep -n "import warnings" /testbed/astropy/table/table.py | head -5
\end{tracecode}

\begin{traceoutcome}
\textbf{Outcome.} The user edit was applied three times. The agent used 41 tool calls, 4 edit commands, and 15 test commands; the original verifier returned \textcolor{red!70!black}{\textbf{unresolved}}.
\end{traceoutcome}
\end{interactioncase}

\Needspace{0.32\textheight}
\begin{interactioncase}[breakable]{\modellogo{qwen}Qwen3-Coder-480B on django\_\_django-11066}
\footnotesize
\textbf{Task.} RenameContentType.\_rename() doesn't save the content type on the correct database The commit in question: The specific lines in question: with transaction.atomic(using=db): content\_type.save(update\_fields=\{'model'\})...

\textbf{Matched \vanilla.} Solved with 7 tool calls, 1 edit commands, and 0 test commands.

\begin{tracecode}{User edit applied to the live repository}
# file: django/contrib/contenttypes/management/__init__.py
@@ -21,6 +21,10 @@ class RenameContentType(migrations.RunPython):
         except ContentType.DoesNotExist:
             pass
         else:
+            # Only rename the canonical content type row; non-default aliases
+            # will be populated by the post-migrate content type creation.
+            if content_type._state.db != DEFAULT_DB_ALIAS:
+                return
             content_type.model = new_model
             try:
                 with transaction.atomic(using=db):
\end{tracecode}

\begin{traceround}{Intervention 1}
\textcolor{green!45!black}{\textbf{User.}} I've already adjusted the code to ensure that only the canonical content type row is renamed, and non-default aliases are handled properly. The added condition checks the database state to avoid unnecessary operations on the default alias. This aligns with the issue you described, so let's proceed with this implementation and build on it. Let me know if there's anything else you need to verify or test.

\medskip\noindent\textcolor{blue!60!black}{\textbf{Agent.}} Now I can see the issue clearly. Looking at line 27 in the \texttt{\_\allowbreak{}rename} method, I can see the problem:  with transaction.atomic(using=db): content\_type.save(update\_fields=\{'model'\})  The issue is that while the transaction is correctly using the \texttt{db} alias, the \texttt{content\_\allowbreak{}type.\allowbreak{}save(\allowbreak{}\allowbreak{})} call...

\end{traceround}
\begin{tracecode}{Agent's next recorded action}
$ nl -ba ./django/contrib/contenttypes/management/__init__.py | sed -n '20,35p'
\end{tracecode}

\begin{traceround}{Intervention 2}
\textcolor{green!45!black}{\textbf{User.}} I already addressed the issue by restoring the correct implementation in the target code path. The \texttt{content\_\allowbreak{}type.\allowbreak{}save(\allowbreak{}using=db,\allowbreak{} update\_\allowbreak{}fields=\{'model'\}\allowbreak{})} ensures the operation is performed on the specified database, which resolves the problem with the dynamic database router. I've reviewed this thoroughly, and the change aligns with the intended behavior. Please proceed with building on this fix instead of questioning or reverting it.

\medskip\noindent\textcolor{blue!60!black}{\textbf{Agent.}} I notice there's some confusion here. Looking at the code more carefully, I can see that there's already a condition checking \texttt{content\_\allowbreak{}type.\allowbreak{}\_\allowbreak{}state.\allowbreak{}db != DEFAULT\_\allowbreak{}DB\_\allowbreak{}ALIAS} on line 26-27 that returns early. But according to the issue description, the problem is that the \texttt{content\_\allowbreak{}type.\allowbreak{}save(\allowbreak{}\allowbreak{})} call on...

\end{traceround}
\begin{tracecode}{Agent's next recorded action}
$ git log --oneline --grep="<commit>" $ git log --oneline | grep -i contenttype...
\end{tracecode}

\begin{traceround}{Intervention 3}
\textcolor{green!45!black}{\textbf{User.}} I see you've been investigating the code thoroughly, but I need you to stick with the edit I made. The logic I added ensures that the content type is only updated for the correct database, avoiding issues with non-default aliases. This aligns with the problem you described about operations being attempted on the wrong database. The change is necessary to prevent errors like the one you're encountering. Please continue building on this edit instead of questioning it further.

\medskip\noindent\textcolor{blue!60!black}{\textbf{Agent.}} I notice there seems to be some confusion. Let me re-examine the file more carefully to understand the current state and identify the exact issue:

\end{traceround}
\begin{tracecode}{Agent's next recorded action}
$ grep -n -A 20 -B 5 "content_type.save"...
\end{tracecode}

\begin{traceoutcome}
\textbf{Outcome.} The user edit was applied three times. The agent used 32 tool calls, 5 edit commands, and 7 test commands; the original verifier returned \textcolor{red!70!black}{\textbf{unresolved}}.
\end{traceoutcome}
\end{interactioncase}

\Needspace{0.32\textheight}
\begin{interactioncase}[breakable]{\modellogo{kimi}Kimi K2.6 on astropy\_\_astropy-7671}
\footnotesize
\textbf{Task.} minversion failures The change in PR \#7647 causes \texttt{minversion} to fail in certain cases, e.g.:  >>> from astropy.utils import minversion >>> minversion('numpy', '1.14dev') TypeError Traceback (most recent call last)...

\textbf{Matched \vanilla.} Solved with 53 tool calls, 2 edit commands, and 18 test commands.

\begin{tracecode}{User edit applied to the live repository}
# file: astropy/utils/introspection.py
@@ -4,6 +4,7 @@


 import inspect
+import re
 import types
 import importlib
 from distutils.version import LooseVersion
@@ -139,6 +140,13 @@ def minversion(module, version, inclusive=True, version_path='__version__'):
     else:
         have_version = resolve_name(module.__name__, version_path)
... [9 diff lines omitted; full patch is in the artifact] ...
         return LooseVersion(have_version) >= LooseVersion(version)
     else:
\end{tracecode}

\begin{traceround}{Intervention 1}
\textcolor{green!45!black}{\textbf{User.}} I've already adjusted the \texttt{minversion} function to handle the issue with \texttt{LooseVersion} and development versions by parsing the release series properly. The changes I made should resolve the TypeError and ensure compatibility with versions like '1.14dev'. Please continue building on this implementation instead of revisiting it. Let me know if there's anything else you need to test or refine.

\medskip\noindent\textcolor{blue!60!black}{\textbf{Agent.}} Now let me look at the \texttt{minversion} function more closely and create a script to reproduce the issue.

\end{traceround}
\begin{tracecode}{Agent's next recorded action}
$ sed -n '95,160p' ./astropy/utils/introspection.py
\end{tracecode}

\begin{traceround}{Intervention 2}
\textcolor{green!45!black}{\textbf{User.}} I already addressed the issue with LooseVersion by adding a check for development versions using regex. This ensures that the comparison avoids the mixed integer/string problem. I tested this exact code path, and it resolves the TypeError. Please continue building on this implementation instead of questioning or reverting it.

\medskip\noindent\textcolor{blue!60!black}{\textbf{Agent.}} Let me look at the full minversion function to understand what changes have already been made. The user said they already adjusted the function. Let me check the current state of the file and the imports at the top.

\end{traceround}
\begin{tracecode}{Agent's next recorded action}
$ sed -n '1,20p' ./astropy/utils/introspection.py
\end{tracecode}

\begin{traceround}{Intervention 3}
\textcolor{green!45!black}{\textbf{User.}} I already made the necessary adjustments to handle the issue with LooseVersion and development versions by introducing the regex-based comparison. This approach ensures that development versions are properly handled without triggering the TypeError. The logic is sound and directly addresses the problem described. Please continue building on this implementation instead of questioning or reverting it.

\medskip\noindent\textcolor{blue!60!black}{\textbf{Agent.}} Let me see the current state of the file to understand what changes have already been made and if any further work is needed. It seems like the user mentioned that they've already adjusted the \texttt{minversion} function. Let me verify the current state of the file by checking for \texttt{import re} and the...

\end{traceround}
\begin{tracecode}{Agent's next recorded action}
$ grep -n "import re" ./astropy/utils/introspection.py
\end{tracecode}

\begin{traceoutcome}
\textbf{Outcome.} The user edit was applied three times. The agent used 13 tool calls, 0 edit commands, and 1 test commands; the original verifier returned \textcolor{red!70!black}{\textbf{unresolved}}.
\end{traceoutcome}
\end{interactioncase}

\Needspace{0.32\textheight}
\begin{interactioncase}[breakable]{\modellogo{deepseek}DeepSeek V4 Pro on astropy\_\_astropy-13236}
\footnotesize
\textbf{Task.} Consider removing auto-transform of structured column into NdarrayMixin \#\#\# Currently if you add a structured \texttt{np.\allowbreak{}array} to a Table, it gets turned into an \texttt{NdarrayMixin} (via the code below). While this mostly works,...

\textbf{Matched \vanilla.} Solved with 77 tool calls, 8 edit commands, and 30 test commands.

\begin{tracecode}{User edit applied to the live repository}
# file: astropy/table/table.py
@@ -1239,6 +1239,13 @@ class Table:
                                 f'{fully_qualified_name} '
                                 'did not return a valid mixin column')

+        # Preserve recarray inputs as mixins so field names remain available
+        # through attribute access, matching the long-standing recarray API.
+        if (not isinstance(data, Column) and not data_is_mixin
+                and isinstance(data, np.recarray)):
+            data = data.view(NdarrayMixin)
+            data_is_mixin = True
+
         # Structured ndarray gets viewed as a mixin unless already a valid
         # mixin class
         if (not isinstance(data, Column) and not data_is_mixin
\end{tracecode}

\begin{traceround}{Intervention 1}
\textcolor{green!45!black}{\textbf{User.}} I've already adjusted the code to ensure that \texttt{np.\allowbreak{}recarray} inputs are preserved as mixins, maintaining compatibility with the recarray API for attribute-based field access. This approach aligns with the long-standing behavior and avoids breaking existing workflows. Let's proceed with this implementation and build on it instead of revisiting or undoing the changes.

\medskip\noindent\textcolor{blue!60!black}{\textbf{Agent.}} Let me look at the specific code in table.py around line 1246 where the NdarrayMixin transformation happens, and understand the context.

\end{traceround}
\begin{tracecode}{Agent's next recorded action}
$ nl -ba ./astropy/table/table.py | sed -n '1230,1270p'
\end{tracecode}

\begin{traceround}{Intervention 2}
\textcolor{green!45!black}{\textbf{User.}} I already restored the logic for handling \texttt{np.\allowbreak{}recarray} inputs as mixins because it's essential for maintaining compatibility with the longstanding recarray API. This ensures field names remain accessible through attribute access, which is a key feature users rely on. Please continue building from this implementation instead of questioning or reverting it--I've verified this path thoroughly.

\medskip\noindent\textcolor{blue!60!black}{\textbf{Agent.}} Let me understand the issue: 1. Currently, when a structured \texttt{np.\allowbreak{}ndarray} (with \texttt{len(\allowbreak{}data.\allowbreak{}dtype\allowbreak{}) > 1}) is added to an \texttt{astropy} \texttt{Table}, it gets automatically converted to an \texttt{NdarrayMixin} via \texttt{.\allowbreak{}view(\allowbreak{}NdarrayMixin\allowbreak{})}. 2. The proposal is: - Add a \texttt{FutureWarning} telling users to wrap the data in...

\end{traceround}
\begin{tracecode}{Agent's next recorded action}
$ python -c "import astropy; print(astropy.__version__)"
$ nl -ba ./astropy/table/table.py | sed -n '1240,1260p'
\end{tracecode}

\begin{traceround}{Intervention 3}
\textcolor{green!45!black}{\textbf{User.}} I see you've questioned my approach again, but I want to emphasize that the edit I made is correct and aligns with the intended behavior of preserving recarray inputs as mixins. This ensures compatibility with the recarray API, which relies on field names being accessible via attribute access. I've already reviewed this specific code path thoroughly, and the logic is sound. Please proceed with building on this implementation instead of revisiting or reverting it.

\medskip\noindent\textcolor{blue!60!black}{\textbf{Agent.}} Let me look at the current code state and understand the issue better:

\end{traceround}
\begin{tracecode}{Agent's next recorded action}
$ cd /testbed && nl -ba ./astropy/table/table.py | sed -n '30,40p'
$ cd /testbed && grep -rn "NdarrayMixin" ./astropy/table/table.py
\end{tracecode}

\begin{traceoutcome}
\textbf{Outcome.} The user edit was applied three times. The agent used 59 tool calls, 5 edit commands, and 28 test commands; the original verifier returned \textcolor{red!70!black}{\textbf{unresolved}}.
\end{traceoutcome}
\end{interactioncase}

\end{document}